\documentclass[aps,pra,10pt,twocolumn,floatfix]{revtex4-2}

\usepackage[utf8]{inputenc}
\usepackage{times}
\usepackage{graphicx}
\usepackage{subfigure}
\usepackage{dcolumn}
\usepackage{bm}
\usepackage{xcolor}
\usepackage{amsmath,amssymb}
\usepackage{verbatim}
\usepackage{mathtools}
\usepackage{epstopdf}
\usepackage{bbold}
\usepackage{verbatim}

\usepackage{hyperref}
\hypersetup{
	colorlinks=true,
	linkcolor=blue,           
	citecolor=blue,           
	filecolor=magenta,        
	urlcolor=cyan
}

\DeclarePairedDelimiter\bra{\langle}{\rvert}
\DeclarePairedDelimiter\ket{\lvert}{\rangle}
\DeclarePairedDelimiterX\braket[2]{\langle}{\rangle}{#1 \delimsize\vert #2}
\DeclarePairedDelimiterX\ketbra[2]{\lvert}{\rvert}{#1 \rangle\hspace{-.25em}\langle #2}

\begin{document}

    \title{Unifying framework for non-Hermitian and Hermitian topology in driven-dissipative systems}

	\author{Clara C. Wanjura}
	\address{Max Planck Institute for the Science of Light, Staudtstraße 2, 91058 Erlangen, Germany}

	\author{Andreas Nunnenkamp}
	\address{Faculty of Physics, University of Vienna, Boltzmanngasse 5, 1090 Vienna, Austria}
	
	\date{\today}
	
	\begin{abstract}
        Recently, a one-to-one correspondence between non-trivial non-Hermitian topology and directional amplification has been demonstrated, theoretically and experimentally, for the case of one complex band.
        Here, we extend our framework to multiple bands and higher spatial dimension. This proves to be far from trivial.
        Building on the singular value decomposition, we introduce a new quantity that we dub \emph{generalised singular spectrum} (GSS). The GSS allows us to define physically meaningful bands related to the system's scattering behaviour and to define invariants for novel notions of point gaps (non-Hermitian topology) and line gaps (Hermitian--like topology), respectively.
        For both invariants, we prove a bulk-boundary correspondence and show that they give rise to two different kinds of topological edge modes.
        We illustrate our results with a 1D non-Hermitian Su-Schrieffer-Heeger (SSH) model and a 2D non-Hermitian model that features corner-to-corner amplification.
        Our work is relevant for many state-of-the-art experimental platforms and it sets the stage for applications such as novel directional amplifiers and non-reciprocal sensors.
	\end{abstract}
	
	\keywords{topology, non-Hermitian topology, dissipation, driven-dissipative quantum systems, reservoir engineering}
	
	\maketitle

\section{Introduction}
Topology is a powerful principle for understanding many physically very different complex systems and has been a prominent research theme in condensed matter physics~\cite{Hasan2010,Bansil2016}.
More recently, a notion of topology in systems experiencing gain and loss has been investigated, sparking the rapidly growing field of non-Hermitian topology~\cite{Ashida2020,Bergholtz2021}.
Such non-Hermiticity can result in a number of remarkable phenomena with no counterpart in closed, Hermitian systems, such as the non-Hermitian skin effect (NHSE)---the localisation of a macroscopic number of eigenvectors at one system edge.
The NHSE not only complicates the study of any residual Hermitian phenomena in the non-Hermitian setting,
it also results in significant theoretical challenges, such as non-orthogonal eigenvectors and the breakdown of the bulk-boundary correspondence~\cite{Xiong2018why}.
This has so far complicated a study of non-Hermitian and Hermitian topology alongside each other, although approaches exist to detect topological boundary states, e.g., via the bi-orthogonal polarisation~\cite{Kunst2018BulkBoundary,Edvardsson2019,Edvardsson2020},
and non-Bloch band theory~\cite{Yao2018edge, Yao2018Chern, Yokomizo2019,Yang2020,Kawabata2020Symplectic,Xiao2020nonhermitian}.

Instead, a framework tailored to the scattering matrix or Green's function has proven successful in the context of point-gap topology: 
Non-trivial point-gap topology has been associated with the phenomenon of directional amplification~\cite{Porras2019,Ramos2021Topological,Wanjura2020,Wanjura2021,Brunelli2023Restoration,McDonald2018,McDonald2020}
which has also been demonstrated in experiments~\cite{Slim2024Optomechanical,Busnaina2024Quantum}.
Non-Hermitian topology is also a resource for sensing~\cite{McDonald2020,Koch2022,Slim2024Optomechanical,konye2024nonhermitian}.
In that context, the singular value decomposition~\cite{Miller2019Lasers} is a powerful tool as it (i)~allows to restore the bulk-boundary correspondence akin to Hermitian systems~\cite{Porras2019,Herviou2019,Ramos2021Topological,Brunelli2023Restoration,monkman2024hidden,mardani2024bulk},
and (ii)~has a straightforward connection to the steady state, the scattering matrix and directional amplification~\cite{Wanjura2020,Wanjura2021,Brunelli2023Restoration,Guo2023Singular,vega2025topological}.

Here, we extend this framework to systems of multiple bands and multiple dimensions. We study the interplay between non-Hermitian and Hermitian topology and, in particular, with our framework reveal phenomena related to remnants of Hermitian topology that were previously overshadowed by the NHSE. Concretely, we build on the singular value decomposition to construct the \emph{generalised singular spectrum} (GSS). The GSS generalises the notion of the eigendecomposition and the notion of bands in Hermitian systems to non-Hermitian systems. It allows us to seamlessly move between Hermitian and non-Hermitian systems and between periodic and open boundary conditions. The GSS also allows to assign different topological invariants to point gaps (non-Hermitian topology) and line gaps (Hermitian topology).

First, we introduce the physical setup of the driven-dissipative systems we will consider throughout this work and introduce the scattering matrix, Green's function and their connection to the singular value decomposition.
Next, we focus on one-dimensional systems and discuss the example of the non-Hermitian Su-Schrieffer-Heeger (SSH) model. We reveal the interplay between Hermitian and non-Hermitian topology which becomes visible with the singular value decomposition. This sets the stage for a more detailed discussion of the non-Hermitian invariant, which we show can be computed from the left and right singular vectors. Furthermore, we define physically meaningful bands which are again related to the behaviour of response functions and we show that we can assign individual winding numbers to each band, which add up to the total winding number of the non-Hermitian system.
This analysis provides a pathway for introducing a new quantity based on the SVD---that we dub generalised singular spectrum (GSS)---which in the Hermitian (and normal) limit recovers the eigenvalues and in the non-Hermitian case generalises the spectrum such that it does not suffer from the NHSE.
The GSS not only gives us access to the point-gap topology, but also allows us to define physically meaningful line gaps.
We assign topological invariants to individual point gaps and line gaps and show that each has a respective bulk-boundary correspondence.
Finally, we demonstrate the generality of our approach by examining a 2D non-Hermitian array with the GSS and reveal corner modes that result in corner-to-corner directional amplification.

Our work is relevant for a large range of experimental platforms including cavity optomechanics~\cite{Slim2024Optomechanical}, superconducting circuits~\cite{youssefi2022topological,Busnaina2024Quantum}, BEC lattices~\cite{Erglis2025Time}, plasmonic waveguides~\cite{Wetter2023Observation}, magnonic systems~\cite{Duine2018Synthetic,Hurst2022Non} and topolectric circuits~\cite{Lee2018}.

\section{Setup}
We consider a collection of bosonic modes, e.g. a cavity array, evolving according to the master equation
$\dot \rho = - \mathrm{i} [\mathcal{H},\rho] + \sum_j \mathcal{D}[L_j] \rho$
with Hamiltonian $\mathcal{H}$, dissipators $\mathcal{D}[L_j]\rho \equiv L_j\rho L_j^\dagger-\frac{1}{2} \{L_j^\dagger L_j, \rho\}$ and jump operators $L_j$. Coherent and dissipative processes are engineered to give rise to an effectively non-Hermitian model~\cite{Wanjura2020,Brunelli2023Restoration,Porras2019,Flynn2020,Flynn2021} (see Methods for more details). In particular, the equations of motions for the bosonic mean fields $\langle a_j\rangle$ can be cast into the form
\begin{align}\label{eq:eoms}
    \langle \dot a_j \rangle & = -\mathrm{i} \sum_{\ell} H_{j,\ell} \langle a_\ell\rangle - \sqrt{\gamma} \langle a_{j,\mathrm{in}}(t)\rangle
\end{align}
with $H$ the non-Hermitian dynamic matrix. $\langle a_{j,\mathrm{in}}(t)\rangle \equiv \langle a_{j,\mathrm{in}}(\omega)\rangle e^{-\mathrm{i}\omega t}$ denotes the input signal at frequency $\omega$ which we inject to probe the system response.
Fig.~\ref{fig:interplay}~\textbf{a} illustrates an example of such a one-dimensional non-Hermitian chain.
The resulting output signals 
$\mathbf{a}_\mathrm{out}\equiv(a_{1,\mathrm{out}}, \dots, a_{N,\mathrm{out}})$ are related to the input signals $\mathbf{a}_\mathrm{in}\equiv(a_{1,\mathrm{in}}, \dots, a_{N,\mathrm{in}})$ and the internal fields $a_j$ via input-output conditions~\cite{Gardiner1985,Clerk2010}, $\langle a_{j,\mathrm{out}}\rangle = \langle a_{j,\mathrm{in}}\rangle + \sqrt{\gamma} \langle a_j\rangle$.
For simplicity, we omitted the frequency argument, $a_j = a_j(\omega)$.
Together with the equations of motion~\eqref{eq:eoms}, this gives rise to the scattering matrix $S(\omega)$ connecting input and output fields $\mathbf{a}_\mathrm{out} = S(\omega) \mathbf{a}_\mathrm{in}$
\begin{align}\label{eq:scattMat}
    S(\omega) & = \mathbb{1} - \mathrm{i} \sqrt{\gamma}(\omega \mathbb{1} - H)^{-1} \sqrt{\gamma}
    = \mathbb{1} - \mathrm{i} \sqrt{\gamma}G(\omega) \sqrt{\gamma}.
\end{align}
The matrix $\gamma\equiv\mathrm{diag}\,(\gamma_1,\dots,\gamma_N)$ contains the coupling rates between the modes and the input-output channels, and $G(\omega)$ denotes the Green's function.
It previously has been shown~\cite{Wanjura2020,Wanjura2021} that in one-dimensional chains with one band non-trivial, non-Hermitian topology leads to directional end-to-end amplification ($\lvert S_{j,\ell}\rvert \neq \lvert S_{\ell,j}\rvert$ and some $\lvert S_{j,\ell}\rvert^2 \geq 1$) with the end-to-end gain $\lvert S_{1,N}\rvert^2$ growing exponentially with the system size (if the directionality of the system is reversed, then, without loss of generality, the end-to-end gain is given by $\lvert S_{N,1}\rvert^2$).

This behaviour can also be understood from the singular value decomposition (SVD)~\cite{Herviou2019,Porras2019,Brunelli2023Restoration} of $H$, for which a non-trivial, non-Hermitian topology is associated with zero modes which are channels of directional amplification (Methods). The SVD decomposes the non-Hermitian dynamic matrix $H = U \Sigma V^\dagger$ into two unitaries $U\equiv(\ket{u_1},\dots,\ket{u_N})$, $V\equiv(\ket{v_1},\dots,\ket{v_N})$ with $\ket{u_j}$ and $\ket{v_j}$ the left and right singular vectors, respectively, and a diagonal matrix $\Sigma\equiv\mathrm{diag}(\sigma_1,\dots,\sigma_N)$ collecting the corresponding non-negative singular values.
Importantly, the SVD can be computed as eigenvalues and vectors of the doubled, Hermitian matrix
\begin{align}\label{eq:SVDdoubled}
    \begin{pmatrix}
        0 & H^\dagger \\ H & 0
    \end{pmatrix}
    \begin{pmatrix}
        \ket{v_j} \\ \ket{u_j}
    \end{pmatrix}
    = \sigma_j \begin{pmatrix}
        \ket{v_j} \\ \ket{u_j}
    \end{pmatrix}.
\end{align}
The SVD is an important analysis tool for non-normal matrices since, in contrast to the eigendecomposition it does not suffer from the non-Hermitian skin effect.
Physically, the singular value decomposition is associated with the scattering response and the steady state, while the eigenvalues are associated with the system dynamics and stability (Methods).

\section{Example: non-Hermitian SSH model}

\begin{figure*}[htbp]       \includegraphics[width=\textwidth]{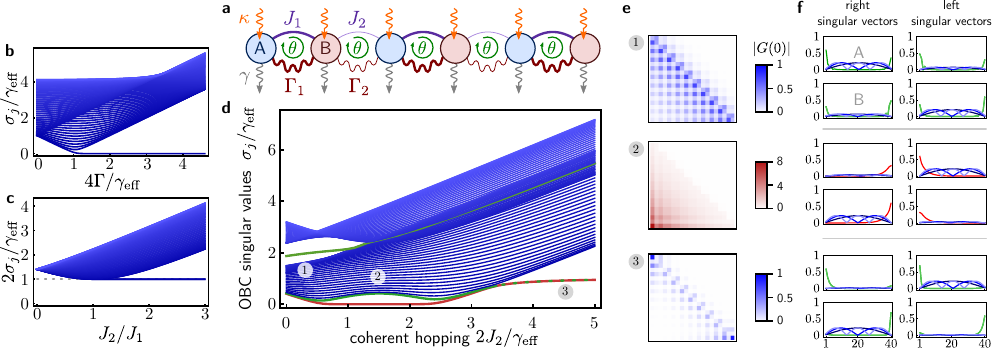}
    \caption{\textbf{Interplay between Hermitian and non-Hermitian topology.}
    \textbf{a}~We consider a non-Hermitian SSH model~\eqref{eq:BlochMatNHSSH} which can be engineered via staggered hoppings $J_1$ and $J_2$ and collective dissipation at rates $\Gamma_1$ and $\Gamma_2$. A phase $\theta$ in the collective dissipator can give rise to non-reciprocity. In addition, balancing local losses $\gamma$ and local incoherent pump rates $\kappa$ can give rise to an amplifying regime.
    \textbf{b}~Without staggering, $J_1=J_2$, $\Gamma_1=\Gamma_2$, the chain resembles a Hatano-Nelson model with local decay that displays directional amplification in non-trivial topological phases~\cite{Wanjura2020}.
    The onset of the topologically non-trivial regime is signalled by the appearance of an exponentially small singular value (singular zero modes) corresponding to an exponentially large gain that appears as we sweep the ratio between collective and local dissipation.
    \textbf{c}~Conversely, if we consider a system with $\Gamma_1=\Gamma_2=0$ but $J_1\neq J_2$, we obtain a Hermitian SSH model with additional local dissipation. In this case, two singular values split off from the bulk as $J_2>J_1$ which is the regime that would be considered as topologically non-trivial in a dissipation-less system.
    \textbf{d}~Finally, in a fully non-Hermitian SSH model, $J_1\neq J_2$, $\Gamma_1\neq \Gamma_2$, we obtain parameter regimes with NH singular zero modes (regime~2) as well as such with Hermitian--like edge modes (regimes~1 and 3). For NH singular zero modes (regime~2), left and right singular vectors localise at opposite ends, panel~\textbf{f}, giving rise to directional end-to-end amplification, panel~\textbf{e}. In contrast, in the Hermitian--like cases (regime~1 and~3) the left and right singular vectors localise at the same edge. If the corresponding singular values are smaller than all bulk singular values (regime~3) the Hermitian--like edge modes become visible in the Green's function, leading to a high reflectivity off the two end sites. In other cases, the Hermitian--like edge modes appear mid-gap so they are overshadowed by the bulk (regime~1).
    Here, \textbf{b}~$\Gamma_1=\Gamma_2=\Gamma$, $2J_1/\gamma_\mathrm{eff}=2J_2/\gamma_\mathrm{eff}=2$, $\theta=\frac{\pi}{2}$;
    \textbf{d}~$\Gamma_1=\Gamma_2=0$, $2J_1/\gamma_\mathrm{eff}=1$;
    \textbf{f}~$J_1/\gamma_\mathrm{eff}=1.5$, $\Gamma_1/\gamma_\mathrm{eff}=0.4$, $\theta=\frac{\pi}{2}$, $\Gamma_2/\gamma_\mathrm{eff}=1$, $N=40$;
    \textbf{e}-\textbf{f}~case~1: $J_1/\gamma_\mathrm{eff}=0.2$, case~2: $J_1/\gamma_\mathrm{eff}=1.5$, case~3: $J_1/\gamma_\mathrm{eff}=4.5$.
    }
    \label{fig:interplay}
\end{figure*}%

We consider the model depicted in Fig.~\ref{fig:interplay}~\textbf{a} which is a bi-partite chain of $2N$ bosonic modes coupled with staggered coherent hopping $\sum_j J_1 a_j^\dagger b_j + J_2 b_j^\dagger a_{j+1} + \mathrm{h.c.}$ and dissipative couplings $\sum_j \Gamma_1 \mathcal{D}[a_j + e^{\mathrm{i}\theta_1} b_j]\rho + \Gamma_2 \mathcal{D}[b_j + e^{\mathrm{i}\theta_2} a_{j+1}]\rho$. 
Additionally, the modes are subject to local dissipation at rate $\gamma$ and gain at rate $\kappa$ which we assume is the same on all sites.
The corresponding Bloch dynamic matrix $H(k)$ under periodic boundary conditions, Eq.~\eqref{eq:eoms}, is given by
\begin{widetext}
\begin{align}\label{eq:BlochMatNHSSH}
    H(k) & =
    \begin{pmatrix}
        -\mathrm{i} \frac{\gamma_\mathrm{eff}}{2}
        & J_1 - \mathrm{i} \frac{\Gamma_1}{2} e^{-\mathrm{i}\theta_1} + (J_2 - \mathrm{i} \frac{\Gamma_2}{2} e^{\mathrm{i}\theta_2}) e^{\mathrm{i} k} \\
        J_1 - \mathrm{i} \frac{\Gamma_1}{2} e^{\mathrm{i}\theta_1} + (J_2 - \mathrm{i} \frac{\Gamma_2}{2} e^{-\mathrm{i}\theta_2}) e^{-\mathrm{i} k}
        & -\mathrm{i} \frac{\gamma_\mathrm{eff}}{2}
    \end{pmatrix}
\end{align}
\end{widetext}
with $\gamma_\mathrm{eff}\equiv \gamma + \Gamma_1+\Gamma_2 - \kappa$.
We consider the singular value decomposition, computed according to Eq.~\eqref{eq:SVDdoubled}, in three different cases:
(i)~the one-band limit $J \equiv J_1=J_2$, $\Gamma \equiv \Gamma_1 = \Gamma_2$;
(ii)~the Hermitian limit, $\Gamma_1 = \Gamma_2 = \kappa = 0$ and $J_1 \neq J_2$, but with added local dissipation $\gamma_\mathrm{eff}\equiv \gamma$;
(iii)~the fully NH SSH model $J_1 \neq J_2$, $\Gamma_1 \neq \Gamma_2$.
We note that the SVD for a related model was recently considered in Ref.~\cite{monkman2024hidden}.

(i)~In the first case, we keep all parameters $J$, $\gamma$, $\kappa$ and $\theta$ fixed, sweep $2\Gamma/\gamma_\mathrm{eff}$ and plot the singular values under OBC, see Fig.~\ref{fig:interplay}~\textbf{b}. We recover the known behaviour of a one-band model: as the strength of the dissipative coupling exceeds the local dissipation ($4\Gamma/\gamma_\mathrm{eff}=1$), the system undergoes a NH topological phase transition~\cite{Brunelli2023Restoration} and the gap closes (the singular values touch $\sigma_j=0$ in the limit $N\to\infty$). As the gap reopens, a zero mode $\sigma_j\propto e^{-\alpha N}$ with some $\alpha>0$ appears. The left and right singular vectors associated with this singular value localise on opposite ends (see Ref.~\cite{Brunelli2023Restoration} for further details on this case). This corresponds to the directional end-to-end amplification channel in non-trivial NH phases.

(ii)~In the second case, we fix $\gamma_\mathrm{eff}=\gamma$ and sweep $J_2/J_1$. In this case, the singular values of Eq.~\eqref{eq:BlochMatNHSSH} are straightforward to compute and are given by $\sigma(k) = \sqrt{(\gamma_\mathrm{eff}/2)^2 + E^2(k)}$ with $E(k) = \pm \sqrt{J_1^2+J_2^2+2J_1 J_2 \cos k}$ the eigenvalues of a Hermitian SSH model, so we expect the singular values to be shifted up by the dissipation $\gamma_\mathrm{eff}$. Indeed, this is also what we observe under OBC, Fig.~\ref{fig:interplay}~\textbf{c}. Note that all singular values in Fig.~\ref{fig:interplay}~\textbf{c} are two-fold degenerate.
As $J_2>J_1$ two edge modes modes split off from the bulk, however, these now correspond to singular values $\sigma_j \approx \gamma_\mathrm{eff}/2$. The corresponding singular vectors behave just as the eigenvectors of a Hermitian SSH model would. Left and right singular vectors are the same (up to a phase) and localise on opposite ends for the $A$ and $B$ sub-lattice.
We notice that it is now possible to remove the edge modes without closing a gap which we will discuss in more detail in later sections. This provides a first glimpse into the fate of Hermitian topology in the presence of dissipation.

(iii)~In the third case, we consider the fully NH SSH model, fix $J_1$, $\Gamma_1$, $\Gamma_2$, and $\theta$, and sweep $2J_2/\gamma_\mathrm{eff}$. $\theta=\frac{\pi}{2}$ is chosen for optimal non-reciprocity.
We can identify three regimes according to Fig.~\ref{fig:interplay}~\textbf{d}, based on the system response encoded in the Green's function shown in Fig.~\ref{fig:interplay}~\textbf{e}.
In the first regime for small $4J_2/\gamma_\mathrm{eff}$, the bulk dominates. The response is clearly non-reciprocal, $\lvert G_{j,\ell}\rvert \neq \lvert G_{\ell,j}\rvert$, but no edge is singled out. In principle, there are two-fold degenerate edge modes between the two bands defined by the singular values in Fig.~\ref{fig:interplay}~\textbf{d} with the corresponding singular vectors localised on both $A$ and $B$ sub-lattices, Fig.~\ref{fig:interplay}~\textbf{f}. However, these edge modes are overshadowed by the bulk of the lower lying band which dominate the Green's function.
As we increase $4J_2/\gamma_\mathrm{eff}$ the NH gap closes and we enter the second regime, Fig.~\ref{fig:interplay}~\textbf{d}. As the gap reopens, a NH topological zero mode appears corresponding to a single set of left and right singular vectors localised on opposite ends for both $A$ and $B$ sub-lattice but on the same ends for both sub-lattices, Fig.~\ref{fig:interplay}~\textbf{f}. This is reflected in the directionally amplifying system response, Fig.~\ref{fig:interplay}~\textbf{e}. The behaviour in the regime is analogous to the single-band case studied in Refs.~\cite{Brunelli2023Restoration,Porras2019} and indeed we relate it to a non-trivial non-Hermitian invariant in the next sections.
Note that in this second regime, the two bulk bands also start to overlap and the edge modes we identified in the first regime disappear.
Finally, we close the NH gap again and transition to a third regime in which two degenerate edge modes appear at finite singular value. The corresponding singular vectors behave analogous to the Hermitian SSH model: left and right singular vectors localise at the same end but on opposite ends for $A$ and $B$ sub-lattices.
Since the edge modes are separated from the bulk by a gap, these edge modes also dominate the Green's function, Fig.~\ref{fig:interplay}~\textbf{e}, making the diagonal element of the first and last site dominant. This implies that the steady-state amplitudes $\langle a_1(0)\rangle$ and $\langle b_N(0)\rangle$ are largest. With the appropriate impedance matching (i.e. choosing $\gamma = \gamma_\mathrm{eff}$ in Eq.~\eqref{eq:scattMat} accordingly), the absorption becomes maximal at the two ends of the chain. We also notice that, since the separation is not as clear as in the NH non-trivial regime, the bulk response is still weakly visible in the Green's function.
This last regime can be understood as the limit when $J_2$ is sufficiently large compared to $\Gamma_1$ and $\Gamma_2$ so that the coherent, staggered couplings give rise to SSH--like behaviour.

Importantly, in all three cases, the features described above are \emph{not} visible in the OBC eigenvectors which for the system shown in Fig.~\ref{fig:interplay}~\textbf{d}-\textbf{f} all localise and hence display the NHSE.

\section{Non-Hermitian topological invariant and the notion of bands}
As the next step, we revisit the definition of the topological winding number for one dimensional non-Hermitian systems. First, we compute the dynamic matrix under periodic boundary conditions (PBC) and express the Bloch dynamic matrix in terms of the SVD
\begin{align}\label{eq:PBCSVD}
    H(k) = \sum_j \sigma_j(k) \ketbra{u_j(k)}{v_j(k)}.
\end{align}
The winding number is typically defined from the determinant of $H(k)$~\cite{Gong2018}
\begin{align}\label{eq:windingNumberMultiBand}
    \nu \equiv \frac{1}{2\pi\mathrm{i}} \int_0^{2\pi} \mathrm{d}k\, \partial_k \ln\det H(k).
\end{align}
Rewriting Eq.~\eqref{eq:windingNumberMultiBand} together with Eq.~\eqref{eq:PBCSVD} lets us split this expression into a sum of winding numbers for individual bands (Supplementary Information)
\begin{align}
    \nu & =
    \frac{1}{2\pi\mathrm{i}} \int_0^{2\pi} \mathrm{d}k\, \mathrm{tr}\,\left(\partial_k \ln H\right) \equiv \sum_j \nu_j
\end{align}
with
\begin{align}\label{eq:windingSkewPolarisation}
    \nu_j = \frac{1}{2\pi\mathrm{i}}  \int_0^{2\pi} \mathrm{d}k\,
    \left(
    \bra{u_j(k)} \partial_k \ket{u_j(k)}
    -
    \bra{v_j(k)}\partial_k \ket{v_j(k)}
    \right).
\end{align}
Here, we introduced the winding number $\nu_j$ for each band $j$ given by the left $\ket{u_j(k)}$ and right singular vectors $\ket{v_j(k)}$ of  the singular-value band $\sigma_j(k)$.
Splitting the winding number above into a sum over winding numbers of individual bands is \emph{only} possible because both $\ket{v_j(k)}$ and $\ket{u_j(k)}$ each span an orthonormal basis $\braket{v_j(k)}{v_\ell(k)} = \braket{u_j(k)}{u_\ell(k)} = \delta_{j,\ell}$ (SI) and because the singular values $\sigma_j(k)$ are real and non-negative (and do not wind themselves).
In contrast, the subspaces defined by the eigendecomposition are not necessarily orthogonal since the Bloch dynamic matrix can already be non-normal making it difficult to connect winding numbers defined on individual eigenvalue bands to physically accessible quantities such as the scattering matrix or the steady state.
In particular, this is because sub-spaces defined through the eigendecomposition under PBC may no longer be identifiable under OBC in cases, when $H$ is non-normal. The sub-spaces defined through the SVD under PBC, however, carry over to OBC.
Singular values and vectors therefore allow for a meaningful division into bands (which also carries over into considerations based on the scattering matrix).

This expression~\eqref{eq:windingSkewPolarisation} for the winding number of band $j$ can be recast into the form (SI)
\begin{align}\label{eq:windingGSS}
    \nu_j = \frac{1}{2\pi\mathrm{i}} \int_0^{2\pi}\mathrm{d}k\,\partial_k\mathrm{Arg}\,\braket{u_j}{v_j} = \frac{1}{2\pi\mathrm{i}} \int_0^{2\pi}\mathrm{d}k\,\frac{\partial_k\braket{u_j}{v_j}}{\braket{u_j}{v_j}},
\end{align}
i.e., an integral over the relative phase between left and right singular vector, which is gauge invariant.
Note that this expression for the winding number is connected to the skew polarisation which has been defined for Hermitian models with chiral symmetry~\cite{Mondragon-Shem2014Topological} (class AIII).

\section{generalised singular spectrum}
\begin{figure*}
    \centering
    \includegraphics[width=\textwidth]{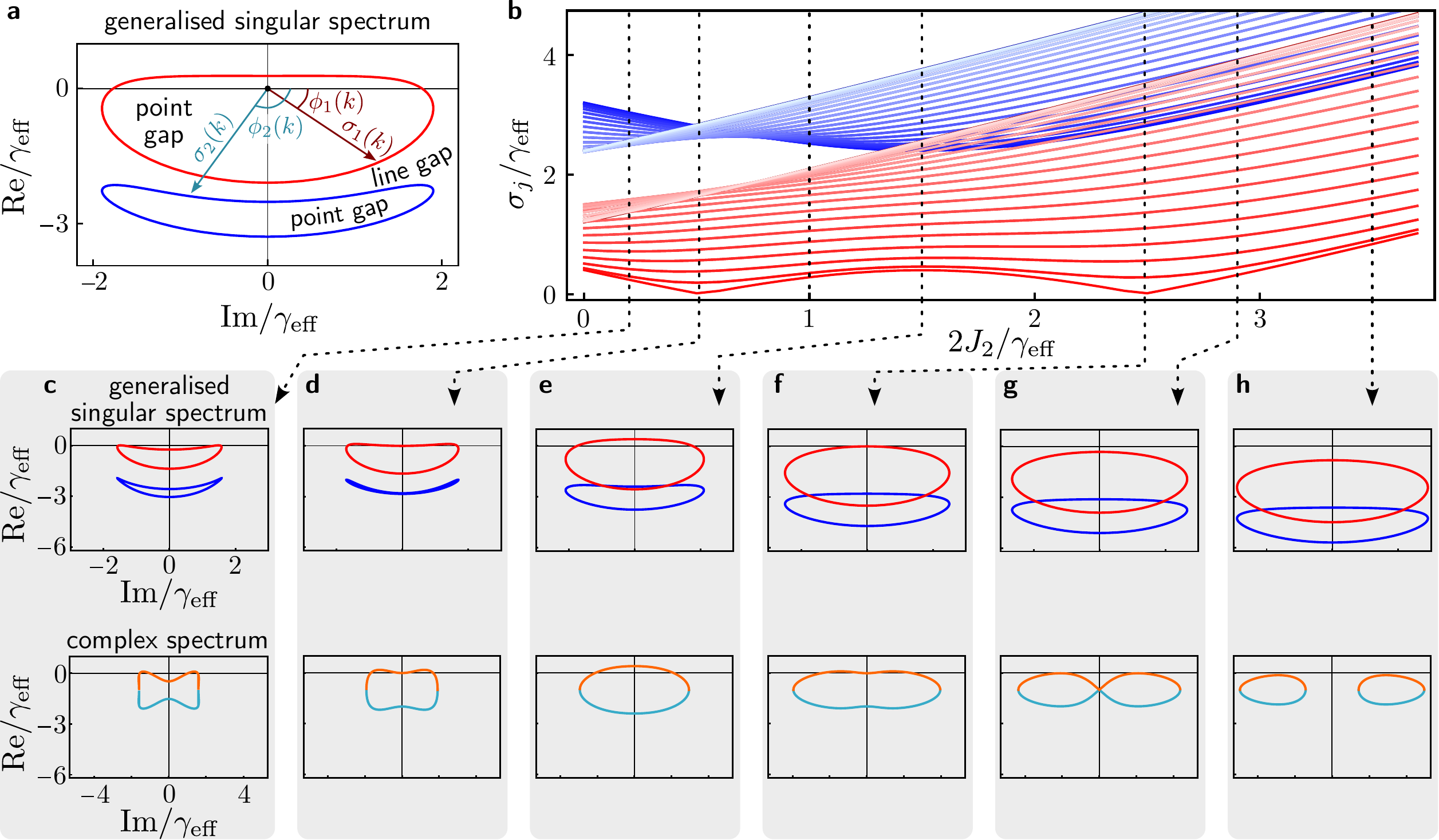}
    \caption{\textbf{The generalised singular spectrum under periodic boundary conditions.}
    We define the generalised singular spectrum (GSS), Eq.~\eqref{eq:generalisedSingularSpectrum}, based on the singular value decomposition which provides a meaningful notion of a complex band for non-Hermitian systems. \textbf{a}~The singular value determines the distance of a point on the complex band from the origin (the pole of the Green's function) and the relative phase between left and right singular vector is the angle of the vector pointing from the origin to a point on the band. Importantly, this angle also determines the winding number of this band, Eq.~\eqref{eq:windingGSS}.
    \textbf{b}~We show the singular values as we sweep $2J_2/\gamma_\mathrm{eff}$ in the NH SSH model, Eq.~\eqref{eq:BlochMatNHSSH}. The corresponding plots of the GSS (top row in panel \textbf{c}-\textbf{h}) reveal the two complex bands which generally differ significantly from the complex spectrum.
    \textbf{c}~We start from an open line gap in the GSS (top row) and neither of the bands encircles the origin. At the same time, the two eigenvalues (bottom row) do not display an open line gap. The different colours denote the two solutions for the GSS and the eigenvalues, respectively. Importantly, we see that the individual eigenvalues are not necessarily periodic modulo $2\pi$, i.e., the individual eigenvalues do not necessarily form a closed loop, whereas each band of the GSS does.
    \textbf{d}~At the NH topological phase transition, one of the GSS bands and one of the eigenvalues touch the origin. The origin is the only point which both GSS and complex spectrum need to have in common, since at this point the determinant vanishes which can both be expressed as product of eigenvalues and as product of the GSS, i.e., $\lvert\mathrm{det}H(k)\rvert = \lvert\Pi_j \varepsilon_j(k)\rvert = \lvert\Pi_j\lambda_j(k)\rvert$ with $\lambda_j$ the $j$th eigenvalue of $H(k)$.
    \textbf{e}~The red GSS band encircles the origin, so this corresponds to a non-trivial NH topological phase. We note that also the line gap in the GSS has closed while it still remains closed in the complex spectrum.
    \textbf{f}~Next, the GSS band touches the origin again signalling a NH topological phase transition.
    \textbf{g}-\textbf{h}~As we further increase $2J_2/\gamma_\mathrm{eff}$, the eigenvalues split into two lobes, while the two eigenvalue solutions still contribute two both lobes (blue and orange colour). At the same time, the GSS does not display any obvious signatures.
    Here, $\Gamma_1/\gamma_\mathrm{eff}=0.4$, $\theta_1=\theta_2=\pi/2$, $2J_1/\gamma_\mathrm{eff}=1.5$, $\Gamma_2/\gamma_\mathrm{eff}=1$, \textbf{b}~$N=40$.}
    \label{fig:GSS}
\end{figure*}

We see from Eq.~\eqref{eq:windingGSS} that the winding number can be computed from the phase
\begin{align}
    \phi_j(k)\equiv \mathrm{Arg}\,\braket{u_j(k)}{v_j(k)}.
\end{align}
This is reminiscent of the one-band case
in which we can simply write $H(k) = \sigma(k) e^{\mathrm{i}\phi(k)}$ with $\phi(k)=\mathrm{Arg}\,H(k)$ and compute the winding number from the phase $\phi(k)$.
Generalising this idea of decomposing our PBC Bloch dynamic matrix into a quantity that serves as measure of distance from the origin and a phase that determines the winding, we can rewrite the singular value decomposition of the Bloch dynamic matrix in the following way
\begin{align}\label{eq:GSSDecomposition}
    H(k) & = \sum_j \sigma_j(k) e^{-\mathrm{i}\phi_j(k)} \ketbra{\tilde u_j(k)}{v_j(k)}
\end{align}
with $\ket{\tilde u_j(k)}=e^{-\mathrm{i}\phi_j(k)} \ket{u_j(k)}$, i.e., the singular vectors remain unchanged up to a phase.
We see that the first term in the sum is now a complex number
\begin{align}\label{eq:generalisedSingularSpectrum}
    \varepsilon_j(k)\equiv\sigma_j(k) e^{-\mathrm{i}\phi_j(k)}
\end{align}
which we call \emph{generalised singular spectrum} (GSS), Fig.~\ref{fig:GSS}~\textbf{a}.
We note that this decomposition can also be calculated under OBC (for further details on its computation, see Methods).
As we now show, the GSS has a number of advantages:
(i)~The GSS allows us to easily define and visualise point gap and line gap topology, Fig.~\ref{fig:GSS}~\textbf{a}; previously, point and line gaps were defined for the spectrum~\cite{Gong2018,Kawabata2018} but not the GSS. We can furthermore associate topological invariants to individual point gaps and line gaps in the GSS.
(ii)~In the limit of normal matrices (which includes the Hermitian case), the GSS \emph{coincides} with the spectrum (eigenenergies). This is because for normal matrices, the SVD can be obtained from the eigendecomposition, namely, by absorbing the phases (the signs in the Hermitian case) of the eigenvalues into the left or right eigenvectors which then become the singular vectors. The GSS reverses this step and generalises it to generic, non-normal matrices. Therefore, the line gaps that appear in the GSS generalise the energy gaps of Hermitian systems to the non-Hermitian domain.
(iii)~One can show a bulk-boundary correspondence for point gaps and line gaps based on the GSS and the SVD (see the next section). In particular, the notion of the bands defined via the SVD or GSS is the same under PBC and OBC.
(iv)~The GSS has a straightforward connection to the scattering matrix which is inherited from the singular value decomposition~\eqref{eq:SVDdoubled} (see Methods).

We now examine the GSS for the NH SSH model, Eq.~\eqref{eq:BlochMatNHSSH}, in Fig.~\ref{fig:GSS} and compare it to the eigendecomposition.
In Fig.~\ref{fig:GSS}~\textbf{b} we show the singular values $\sigma_j$ as we sweep the hopping strength $J_2$.
Note that the PBC singular values, Fig.~\ref{fig:GSS}~\textbf{b}, and OBC singular values, Fig.~\ref{fig:interplay}~\textbf{d}, agree up to the edge modes as expected.
We notice that there are two NH gap closings (i.e., the singular values touch $\sigma_j=0$~\cite{Brunelli2023Restoration}). Furthermore, the band gap between the two bands is open for sufficiently small $J_2$ but then closes and remains closed for larger $J_2$.
In Figs.~\ref{fig:GSS}~\textbf{c} to \textbf{h} we plot the generalised singular spectrum and compare it to the complex spectrum which vastly differs from the generalised singular spectrum.

We note that already individual GSS bands allow for the definition of point gaps in Fig.~\ref{fig:GSS}~\textbf{a},
while the two solutions for the complex eigenvalues shown in the bottom row may not necessarily form a closed loop individually.
Furthermore, we notice that we can define a line gap between the two bands of the GSS which is now defined to exist when the two bands do not touch or cross.

In Fig.~\ref{fig:GSS}~\textbf{c} both GSS bands are trivial, i.e., they do not encircle the origin. However, the line gap between the two bands is open and we recall that under OBC we found four edge modes, Fig.~\ref{fig:interplay}.
As we sweep $J_2$, we see that first one of the GSS bands touches the origin in Fig.~\ref{fig:GSS}~\textbf{d} and one singular value becomes zero, Fig.~\ref{fig:GSS}~\textbf{b}, i.e., the NH gap closes. At the same time, the spectrum passes through the origin which is expected since a zero singular value results in a zero determinant which implies at least one zero eigenvalue.
As the band moves across the origin, the NH gap reopens and the winding number associated with that point gap becomes non-trivial in Fig.~\ref{fig:GSS}~\textbf{e}.
We also note that the line gap closes between Fig.~\ref{fig:GSS}~\textbf{d} and \textbf{e}.
As the line gap in the GSS closes, so does the gap in the singular values (due to chiral symmetry constraints of the doubled matrix which is used to define the SVD and the GSS, this is always the case, see SI).
Under OBC, this meant that the four edge states disappeared, Fig.~\ref{fig:interplay}. Note that the GSS resolves the degeneracies in the singular values in Fig.~\ref{fig:interplay}~\textbf{d} after the line gap has closed.
In Fig.~\ref{fig:GSS}~\textbf{f}, the NH gap closes again as the band crosses the origin and the winding number becomes trivial again in Fig.~\ref{fig:GSS}~\textbf{g}.
At the same time, the spectrum starts to split into two and a line gap has opened in Fig.~\ref{fig:GSS}~\textbf{h} which is not visible in the GSS. We also note that the opening of this line gap in the spectrum is not met by a change in the singular vectors. The edge modes that appeared in Fig.~\ref{fig:interplay} appear only for larger values of $J_2$.
Therefore, line gaps in the spectrum are neither necessarily related to edge modes that appear in the singular vectors nor to the scattering response and the steady state which are determined by those singular vectors.

The GSS highlights an important point: the line gap did not reopen between Fig.~\ref{fig:GSS}~\textbf{g} and Fig.~\ref{fig:GSS}~\textbf{h}, yet we saw SSH--like localised singular vectors appear in Fig.~\ref{fig:interplay}. This is a signal that these edge modes are in fact trivial. In fact, they can be removed without any changes to the line gap simply by increasing the local dissipation (SI). At the same time, the line gap in the spectrum remains open.

\section{Unifying framework for non-Hermitian and Hermitian topology}
\begin{figure}
    \centering
    \includegraphics[width=\linewidth]{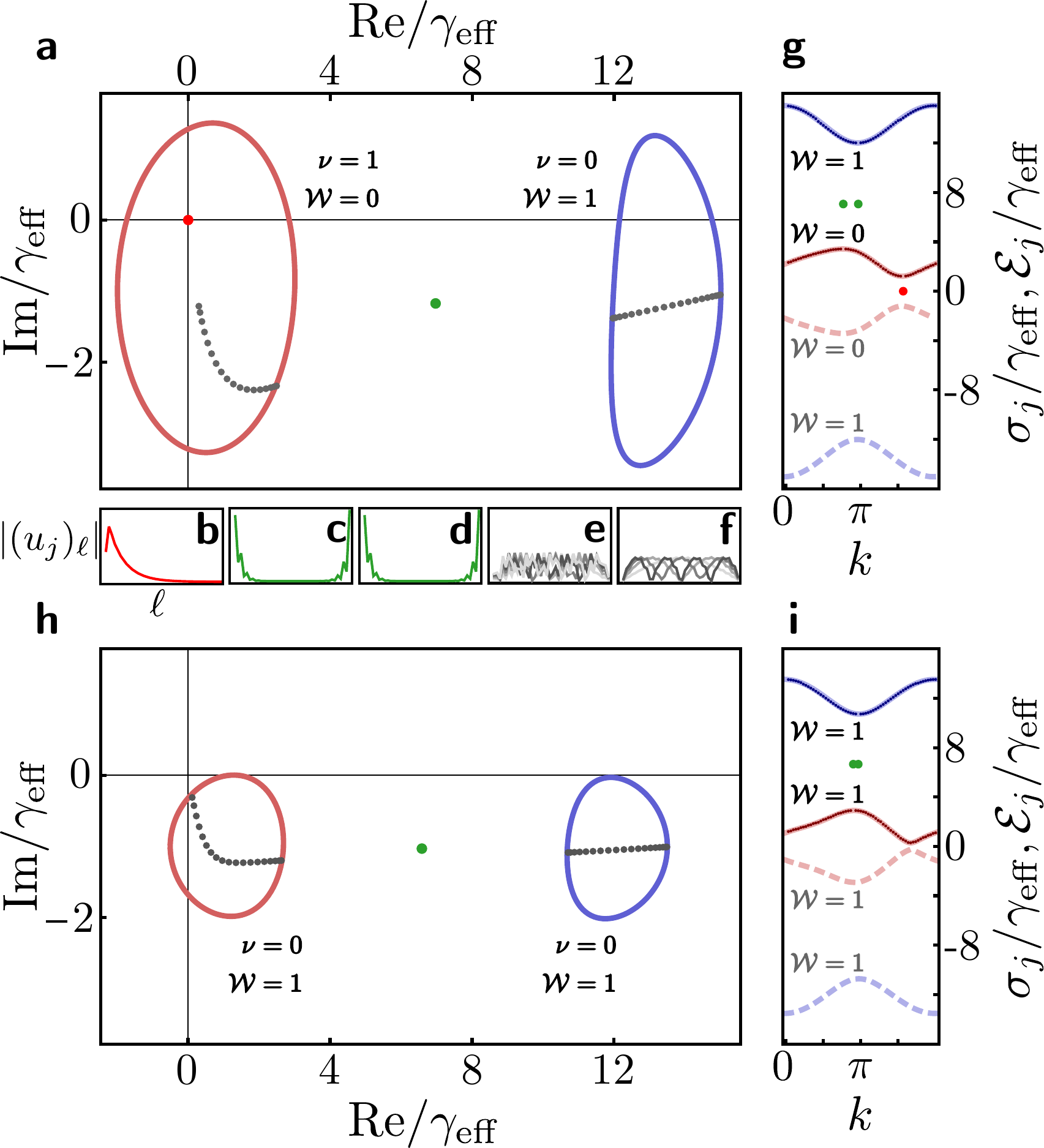}
    \caption{\textbf{Non-Hermitian and Hermitian--like edge modes in the generalised singular spectrum.}
    \textbf{a}, \textbf{h}~Generalised singular spectrum (GSS) under periodic (lines) and open boundary conditions (dots) for the NH SSH model~\eqref{eq:BlochMatNHSSH}. Panels \textbf{g}, \textbf{i} show the corresponding singular value band structure of the doubled Hermitian model~\eqref{eq:SVDdoubled} with the NH bands ($\sigma\geq0$) shown as solid line and the additional Hermitian bands ($\mathcal{E}_j<0$) shown as dashed lines.
    \textbf{a}-\textbf{g} show a case with both NH topological edge modes (panel~\textbf{b}) and Hermitian--like edge modes (panels~\textbf{c} and \textbf{d}) while the bulk modes of the two bands are plane waves (panels~\textbf{e} and \textbf{f}).
    In panels \textbf{h} and \textbf{i} the NH topological edge mode has disappeared as the left-most GSS band has crossed the origin.
    In the doubled Hermitian model, the invariant associated with each gap can be computed by summing over $\mathcal{W}$ modulo $2$ starting from the lowest band up to the band gap, e.g., for the gap at the top in panel \textbf{g} with the two Hermitian--like edge modes, we obtain $(1+0+0)\bmod 2=1$ resulting in two Hermitian--like edge modes.
    In panel \textbf{i}, we obtain for the gap at the top $(1+1+1)\bmod 2=1$ leading again to two Hermitian--like modes.
    This results in the formula given in the main text for the total number of mid-gap edge modes.
    Here, $\Delta/ \gamma_\mathrm{eff} = 6.5$, $\theta_1 = \pi/2$, $2J_1/\gamma_\mathrm{eff} = 1.5$, $\theta_2 = \pi/2$, $2J_1/\gamma_\mathrm{eff} = 5.5$
    \textbf{a}-\textbf{g}~$4\Gamma_1/\gamma_\mathrm{eff} = 4\Gamma_2/\gamma_\mathrm{eff}= 0.78$.
    \textbf{h}-\textbf{i}~$4\Gamma_1/\gamma_\mathrm{eff} = 4\Gamma_2/\gamma_\mathrm{eff}= 1.5$.
    }
    \label{fig:edgeModes}
\end{figure}%
We now proceed to show that we can assign invariants to point gaps and line gaps, respectively, and that each comes with a bulk-boundary correspondence.

First, we show the bulk-boundary correspondence for point gaps. 
The winding number $\nu_j$ introduced in Eq.~\eqref{eq:windingGSS} is assigned to the specific point gap of that GSS band, and the sum of all winding numbers equals the total winding number $\nu=\sum_j \nu_j$ of $\mathrm{det}H(k)$. This winding number is in fact also the relevant topological invariant of one-dimensional systems with chiral symmetry, such as the corresponding doubled matrix $\mathcal{H}(k)$, Eq.~\eqref{eq:SVDdoubled}, from which we compute the SVD. For such Hermitian systems, the following bulk-boundary correspondence has been shown~\cite{ryu2010topological}: under OBC, there are $\lvert\nu\rvert$ pairs of localised eigenvectors in the central gap, typically arising at eigenvalues $\pm e^{-\beta N}$ with some $\beta>0$~\cite{Asboth2016}.
The SVD of the non-Hermitian dynamic matrix is the same as selecting only the non-negative eigenvalues and corresponding eigenvectors of the doubled matrix~\eqref{eq:SVDdoubled} with chiral symmetry. Hence, we inherit the topological classification for Hermitian matrices with chiral symmetry and obtain $\lvert\nu\rvert$ localised singular vectors at exponentially small singular value.
The corresponding left and right singular vectors localise at opposite ends, so the winding numbers determine the number of directional amplification channels.
 
We illustrate the bulk-boundary correspondence for point gaps in Fig.~\ref{fig:edgeModes} for the NH SSH model.
Fig.~\ref{fig:edgeModes}~\textbf{a} and \textbf{h} show the GSS under PBC and OBC. The winding number as point gap invariant can be read off directly for each band. When the winding number (w.r.t.~the origin) is non-zero, one singular zero mode appears in the point gap corresponding to $\nu=+1$. The associated singular vector is localised at the edge, Fig.~\ref{fig:edgeModes}~\textbf{b}.
Note, that the left singular vector in Fig.~\ref{fig:edgeModes}~\textbf{b} corresponding to the non-Hermitian--like edge mode is localised on site $2$ instead of $1$, while the Hermitian like edge modes localise on site $1$. This implies that the largest directional gain is now obtained between site $2$ and $N$. 
The localisation of Hermitian--like and non-Hermitian--like edge modes on different sites ensures that the singular vectors are orthogonal among each other (for both right and left singular vectors, respectively).

Second, we turn our attention to a bulk-boundary correspondence for line gaps in the GSS. In order to protect edge modes within other gaps than at zero, we require other, additional symmetries that do not necessitate pairs of zero modes to appear at positive and negative eigenvalues.
Specifically, we show that when in addition to chiral symmetry, the doubled matrix $\mathcal{H}(k)$, Eq.~\eqref{eq:SVDdoubled}, also possesses particle-hole symmetry (PHS), i.e., there exists a unitary such that $U^\dagger \mathcal{H}^*(k) U = -\mathcal{H}(-k)$ (and with this, $\mathcal{H}(k)$ also automatically possesses time-reversal symmetry),
we can compute a line-gap invariant based on the Zak phase of each individual band, and this invariant determines the number of mid-gap states under OBC within this line gap.
It is straightforward to see from Eq.~\eqref{eq:BlochMatNHSSH} that for $\theta_j=\pm\frac{\pi}{2}$, $\mathcal{H}(k)$ has PHS symmetry (Methods).
We define an integer-quantised invariant $\mathcal{W}_j$ that takes inspiration from the Zak phase of the bands of the doubled matrix $\mathcal{H}(k)$
\begin{align}\label{eq:ZakPhase}
    \mathcal{W}_j = \frac{1}{\pi\mathrm{i}} \int_0^{2\pi} \mathrm{d}k\, \big(&
    \bra{u_j(k)} \partial_k \ket{u_j(k)} \notag \\
    +
    &\bra{v_j(k)}\partial_k \ket{v_j(k)}
    \big) \bmod 2.
\end{align}
In electronic systems, the Zak phase is proportional to the polarisation, i.e., by how much a charge is shifted within a unit cell.
Note that the sum $\mathcal{W} = \sum_j \mathcal{W}_j$ is related to the sum of the winding numbers $\nu=\sum_j\nu_j$ according to
\begin{align}
    \mathcal{W} = \nu \bmod 2.
\end{align}
Making use of the analogy between the SVD of a non-Hermitian matrix and the eigendecomposition of the Hermitian doubled matrix~\eqref{eq:SVDdoubled}, we can draw on known results for Hermitian systems with PHS symmetry in one dimension (class D).
In particular, PHS symmetry of $\mathcal{H}(k)$ ensures that $\mathcal{W}_j$ is $\mathbb{Z}_2$ quantised and can only change due to a closure of the gap between different SVD bands, i.e., through line-gap closures.
Furthermore, in Hermitian systems with PHS, the following bulk-boundary correspondence holds~\cite{Rhim2017Bulk}: the number of edge states in the band gap under OBC is equal to $2\lvert\sum_j \mathcal{W}_j\rvert$, in which the sum runs over the bands below below the Fermi energy~\cite{King-Smith1993Theory}.
Hence, in the non-Hermitian case, we obtain an analogous bulk-boundary correspondence for the singular value decomposition: we order the bands in increasing order of their singular values and introduce the line-gap invariant
\begin{align}\label{eq:lineGapInvariant}
    \Lambda_\ell=\mathcal{W} + \sum_{j=1}^\ell \mathcal{W}_j \bmod 2
\end{align}
for the line gap between band $\ell$ and $\ell+1$. 
This is analogous to adding polarisations in Hermitian, fermionic systems~\cite{King-Smith1993Theory,Vanderbilt1993Electric} with $\mathcal{W}_j = 2P_j$ twice the polarisation $P_j$ of a band in the analogous fermionic system.
Transferring these results to the non-Hermitian setting, the bulk-boundary correspondence then dictates that under OBC, we recover $2\lvert\Lambda_\ell\rvert$ Hermitian--like singular edge modes inside the band gap, i.e., left and right singular vectors both display the same localisation characteristics: they are are localised on the $A$ sub-lattice at one end and on the $B$ sub-lattice at the other end.

We again illustrate the bulk-boundary correspondence for line gaps in Fig.~\ref{fig:edgeModes} for the NH SSH model. The invariant associated with each gap is the same as of the doubled Hermitian system with eigenvalues $\mathcal{E}_j$ which has bands $\pm \sigma_j(k)$, Fig.~\ref{fig:edgeModes}~\textbf{g}, so we sum over the values of the invariant $\mathcal{W}_j$ below the gap of interest to obtain the line gap invariant~\eqref{eq:lineGapInvariant}. A non-trivial value of the line gap invariant gives rise to two Hermitian-like edge modes within the line gap, Fig.~\ref{fig:edgeModes}~\textbf{a}. The corresponding edge modes are localised at both edges, Figs.~\ref{fig:edgeModes}~\textbf{c} and \textbf{d}. In Figs.~\ref{fig:edgeModes}~\textbf{h} and \textbf{i} the point gap invariant is trivial and the NH singular zero mode has disappeared, but the two Hermitian-like edge modes within the line gap persist.

\section{Corner modes in 2D non-Hermitian systems}
\begin{figure}
    \centering
    \includegraphics[width=0.5\textwidth]{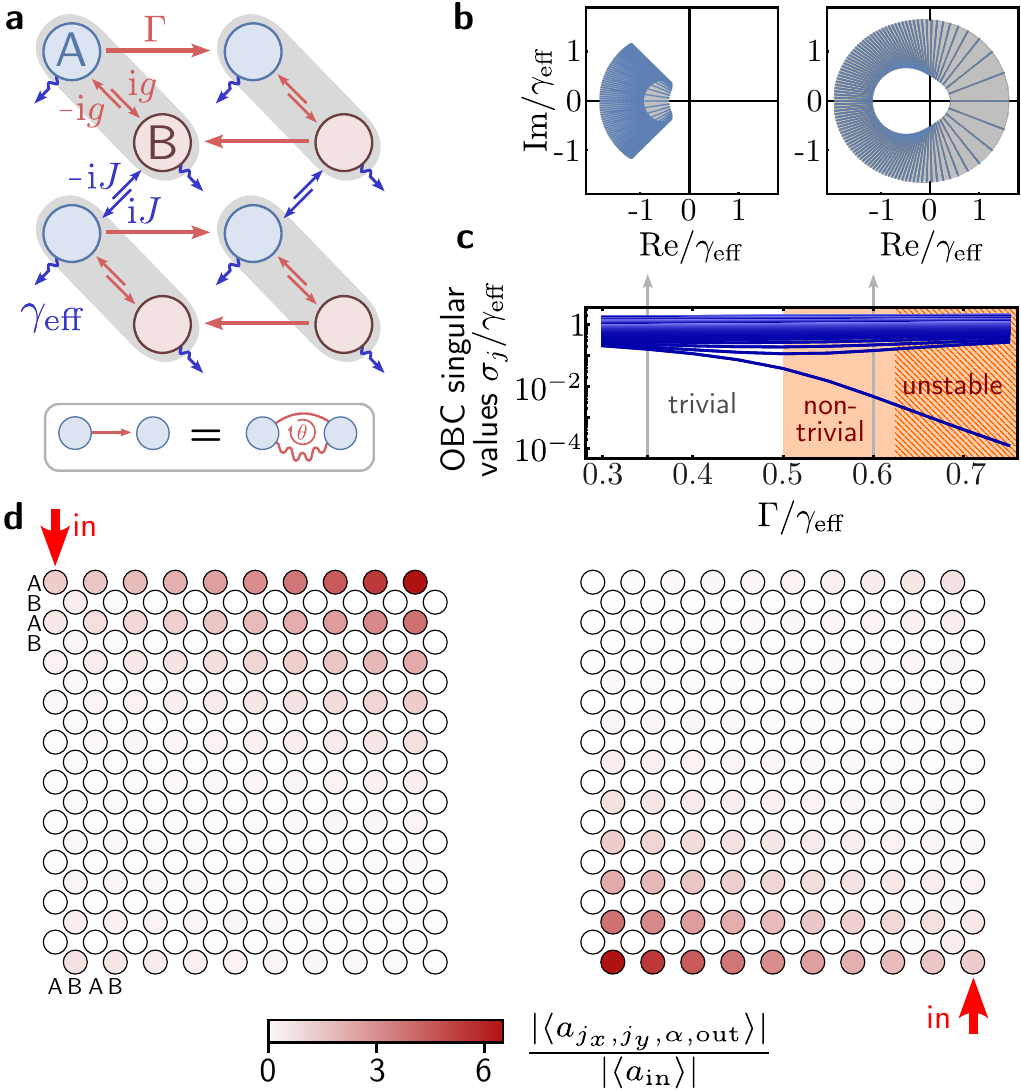}
    \caption{\textbf{Corner modes in a non-Hermitian model.}
    \textbf{a}~Sketch of the 2D lattice giving rise to corner modes for $2\times2$ unit cells (grey areas). Perfectly unidirectional couplings of strength $\Gamma$ connecting $A$ ($B$) sub-lattices of neighbouring unit cells from left to right (right to left) are engineered through interfering hopping and dissipative hopping (inset below). $A$ and $B$ sub-lattices within one unit cell are coupled via imaginary hopping of strength $g$ while adjacent unit cells are coupled vertically via imaginary hopping at strength $J$. Each mode experiences effective local losses at rate $\gamma_\mathrm{eff}$ which is the result of local losses via waveguides, the contributions from the non-local dissipator as well as incoherent pumping to compensate some of the dissipation.
    \textbf{b}~Generalised singular spectrum (GSS) in a trivial regime (left) and in a non-trivial regime (right) at a finite system size of $100\times 100$ unit cells (blue dots) and in the infinite limit (grey area). In the non-trivial regime, the GSS wraps around the origin.
    \textbf{c}~Singular values under open boundary conditions. In the non-trivial regime, two singular values split off the bulk and go to zero in the infinite system limit.
    In the dynamically unstable regime (hashed are), the system is pumped too strongly and some of the imaginary parts of the OBC eigenvalues of the dynamic matrix become positive.
    \textbf{d}~Non-trivial winding in the GSS gives rise to corner modes which yields corner-to-corner directional amplification for which the gain is determined by the inverse of the corresponding small singular values. The red arrow highlights the site at which the input signal is injected.
    Here, $g/\gamma_\mathrm{eff}=0.6$, $J/\gamma_\mathrm{eff}=1$, \textbf{c}-\textbf{d}~system size $10\times 10$ unit cells, \textbf{d}~$\Gamma/\gamma_\mathrm{eff}=0.6$.
    }
    \label{fig:HOTI}
\end{figure}%

We now demonstrate the generality of our approach by applying it to a two-dimensional model.
Specifically, we consider a non-Hermitian model derived from the Hermitian Benalcazar-Bernevig-Hughes (BBH) model introduced in Ref.~\cite{Benalcazar2017Quantized,Benalcazar2017Electric}. In the Hermitian context, it is known for higher-order topology related to corner modes. Here, we consider the model illustrated in Fig.~\ref{fig:HOTI}~\textbf{a} (see Methods for further details). The Bloch matrix derived from the dynamic matrix governing the evolution of the mean fields is given by
\begin{align}\label{eq:modelHOTI}
    H(k_x, k_y) & \equiv
    \begin{pmatrix}
        -\frac{\gamma_\mathrm{eff}}{2} - \Gamma e^{\mathrm{i}k_x} & - g - J e^{\mathrm{i}k_y} \\
        g + J e^{-\mathrm{i}k_y} & -\frac{\gamma_\mathrm{eff}}{2} - \Gamma e^{-\mathrm{i}k_x}
    \end{pmatrix}
\end{align}
with $\gamma_\mathrm{eff} \equiv \gamma - \kappa+2\Gamma$.
A related model has been seen to give rise to a higher-order non-Hermitian skin effect~\cite{Kawabata2020Higher}. Importantly, here, we take into account the contribution of $\Gamma$ to the local dissipation $\gamma_\mathrm{eff}$ which is essential for dynamical stability.
The doubled, Hermitian matrix $\mathcal{H}$ constructed from $H(k)$ according to Eq.~\eqref{eq:SVDdoubled} corresponds exactly to the Hermitian model introduced in Ref.~\cite{Benalcazar2017Quantized,Benalcazar2017Electric}---the Benalcazar-Bernevig-Hughes (BBH) model.
Here, we work in a regime, where the edge polarisation of the doubled Hermitian model~\eqref{eq:SVDdoubled} is non-trivial only along $x$ but not along $y$. We do this to ensure that there is a parameter regime in which the resulting non-Hermitian system is dynamically stable. 

Corner states in the Hermitian model are connected to corner charges computed from the sum of the edge polarisations and the bulk quadrupole moment.
The edge polarisations can be related to a set of winding numbers along $x$ and $y$, respectively, while the quadropole moment is related to the product of the two~\cite{Benalcazar2017Electric}.
Since the eigendecomposition of Eq.~\eqref{eq:SVDdoubled} computes the SVD of our non-Hermitian model, we can formulate an analogous topological characterisation in terms of the SVD for the NH HOTI model.

Specifically, we can characterise the topology of band $j$ for the NH model in terms of a set of winding numbers $\nu_{j,\alpha}$ with $\alpha = x,y$ (the derivation is analogous to the 1D case, see Methods)
\begin{align}
    \nu_{j,\alpha} & \equiv \frac{1}{(2\pi)^2\mathrm{i}} \int_0^{2\pi} \mathrm{d}\mathbf{k} (\bra{u(\mathbf{k})}\partial_{k_\alpha}\ket{u(\mathbf{k})} - \bra{v(\mathbf{k})}\partial_{k_\alpha}\ket{v(\mathbf{k})}).
\end{align}
The total winding number for band $j$ is then given by $\nu_j = \nu_{j,x} + \nu_{j,y}$. Again, we inherit the bulk-boundary correspondence of the Hermitian doubled model~\eqref{eq:SVDdoubled} with chiral symmetry.
Specifically, if the quadrupole moment of the Hermitian doubled model is zero (which it is for the parameters in Fig.~\ref{fig:HOTI}), the total winding number $\sum_j \nu_j$ counts the number of zero singular modes that appear under OBC~\cite{Trifunovic2020Bulk,Ren2021Quadrupole,Yang2023Wannier}.
Just as in the 1D case, the winding numbers can also be computed from the GSS
\begin{align}
    \nu_{j,\alpha}
    & = \frac{1}{(2\pi)^2\mathrm{i}} \int_0^{2\pi}\mathrm{d}\mathbf{k}\,\partial_{k_\alpha}\mathrm{Arg}\,\braket{u_j}{v_j} \notag \\
    & = \frac{1}{(2\pi)^2\mathrm{i}} \int_0^{2\pi}\mathrm{d}\mathbf{k}\,\frac{\partial_{k_\alpha}\braket{u_j}{v_j}}{\braket{u_j}{v_j}}.
    \label{eq:windingGSSHOTI}
\end{align}
We plot the GSS for our NH two-dimensional model~\eqref{eq:modelHOTI} in Fig.~\ref{fig:HOTI}~\textbf{b}. The GSS fills an area which in topologically non-trivial regions with $\nu_{j,\alpha} \neq 0$ encloses the origin, while it does not enclose the origin in trivial regions. The GSS consists of two degenerate bands which overlap perfectly but envelope the origin in opposite directions (Methods).
In the non-trivial case the GSS wraps around the origin for any fixed $k_y$ as we vary $k_x$, while it does not for a fixed $k_x$ as we vary $k_y$. Specifically, $\nu_{j,x}=\pm1$ when $\gamma < \Gamma$ for the two bands, respectively, while $\nu_{j,y}=0$.

Correspondingly, in the non-trivial regime, two zero singular modes (one for each of the degenerate bands corresponding to the top and bottom edge, respectively) appear, Fig.~\ref{fig:HOTI}~\textbf{c}, which give rise to the amplification of an input signal to another corner, Fig.~\ref{fig:HOTI}~\textbf{d}. Fig.~\ref{fig:HOTI}~\textbf{d} shows the dominant response of the system; if an input signal is injected at any other site, the response is weaker (Methods).

This analysis shows that non-Hermitian topological models in higher dimensions can have a hidden point-gap topology that is revealed with the help of the GSS. Analogous to the 1D case, non-trivial topology gives rise to directional corner-to-corner amplification.

\section{Conclusions}
In this work, we developed a framework for the study of non-Hermitian systems with multiple bands. We introduced the generalised singular spectrum
which allowed us to introduce a physically meaningful notion of (complex) bands as it connects to physical properties such as the scattering response. We identified point gaps as well as line gaps in the generalised singular spectrum and assigned topological invariants to both types of gaps. We showed a bulk-boundary correspondence for both point gap invariants (non-Hermitian topology) and line gap invariants (Hermitian--like topology) giving rise to edge modes of different character. In particular, in one dimension, the zero modes associated with non-trivial values of the point gap invariant correspond to directional amplification between opposite ends, while in two dimensions, we obtain corner modes resulting in corner-to-corner directional amplification.
Our framework is relevant for many state-of-the-art experimental platforms,
opening the door to applications such as novel directional amplifiers and sensors based on non-Hermitian or Hermitian topology.

Due to its generality, our framework has furthermore prepared the ground for a number of theoretical studies including a symmetry classification for the generalised singular spectrum and the role of non-Hermitian symmetries~\cite{Gong2018,Kawabata2018}, the impact of disorder~\cite{Kawabata2023Singular}, as well as the systematic exploration of higher dimensional systems, which is currently still technically demanding~\cite{Wang2024Amoeba}. We speculate that the line gap invariant introduced here based on the generalised singular spectrum could also correctly count the topological boundary modes in the eigendecomposition and that robust topological boundary modes can only be removed through line gap closures in the generalised singular spectrum.
The SVD may also provide a natural connection between the Green's function and the generalised Brillouin zone~\cite{Xue2021Simple}.
Beyond that, the duality between non-Hermitian systems and Hermitian systems with chiral symmetry also offers a new perspective on Hermitian systems---especially in the context of higher-order topology which is an active topic of current research.

\section{Acknowledgements}
C.C.W. would like to thank
Jan Behrends,
Wojciech Jankowski,
Flore Kunst,
Maxine McCarthy,
Vittorio Peano,
Hannah Price,
and
Henning Schomerus
for inspiring discussions.
This research was funded in whole or in part by the Austrian Science Fund (FWF) [10.55776/COE1]. For Open Access purposes, the authors have applied a CC BY public copyright license to any author accepted manuscript version arising from this submission.

%


\appendix


\section*{Methods}
\subsection*{Further details about the setup}
In the main text, we consider a typical setup of  coupled bosonic modes evolving according to the master equation
$\dot \rho = - \mathrm{i} [\mathcal{H},\rho] + \sum_j \mathcal{D}[L_j] \rho$ giving rise to effectively non-Hermitian dynamics as described by Eqs.~\eqref{eq:eoms}.
A typical Hamiltonian consists of a system Hamiltonian $\mathcal{H}_\mathrm{sys}$ which encodes couplings between modes, and a Hamiltonian that describes the influence of the input field $a_{j,\mathrm{in}}(\omega)$ at frequency $\omega$ which is employed in typical experiments to probe the system response
\begin{align}\label{eq:coherentTerms}
    \mathcal{H} & = \mathcal{H}_\mathrm{sys} + \mathrm{i} \sum_{j} \sqrt{\gamma_j} a_j^\dagger a_{j,\mathrm{in}}(\omega) e^{-\mathrm{i}\omega t} + \mathrm{h.c.}
\end{align}
For instance, the system Hamiltonian $\mathcal{H}_\mathrm{sys}=\sum_{j,\ell} J_{j,\ell} a_{j}^\dagger a_\ell + \mathrm{h.c.}$
describes hopping between sites $j$ and $\ell$. 
The dissipators $\mathcal{D}[L_j] \rho$ in in the master equation may be local or non-local, e.g., $L_j = \sqrt{\gamma} a_j$ for local dissipation, $L_j = \sqrt{\kappa} a_j^\dagger$ for local gain, and $L_{j,\ell} = \sqrt{\Gamma} (a_j + e^{\mathrm{i}\theta} a_{\ell})$ for dissipative coupling between sites $j$ and $\ell$. The latter can be thought of as hopping via a lossy mode.

The equations of motion~\eqref{eq:eoms} in a frame rotating at the frequency $\omega$ of the drive are generally solved by
\begin{align}\label{eq:dynSol}
    \langle \mathbf{a}(t) \rangle & = e^{-\mathrm{i} (H-\omega) t} \mathbf{c} - \mathrm{i}\sqrt{\gamma} (\omega - H)^{-1} \langle\mathbf{a}_\mathrm{in}(\omega)\rangle
\end{align}
in which we introduced the vectors $\mathbf{a}\equiv(a_1,\dots,a_N)^\mathrm{T}$, $\mathbf{a}_\mathrm{in}\equiv(a_{1,\mathrm{in}},\dots,a_{N,\mathrm{in}})^\mathrm{T}$ and $\mathbf{c} \equiv \langle \mathbf{a}(0)\rangle + \mathrm{i}\sqrt{\gamma} (\omega-H)^{-1} \langle \textbf{a}_\mathrm{in}(\omega)\rangle$.
For sufficiently small times, the first, time-dependent term dominates. Assuming dynamical stability (which is required for such a linear model to be valid), i.e., the imaginary parts of all eigenvalues $\lambda_j$ of $H$ are negative,
the first term in Eq.~\eqref{eq:dynSol} describes the evolution to the steady state. At times $t\gg 1/\mathrm{max} \, \mathrm{Im} \lambda_j$,
the system response to the input field dominates. In particular, the steady state is given by $\langle\mathbf{a}_\mathrm{ss}\rangle = - \mathrm{i}\sqrt{\gamma} (\omega - H)^{-1} \langle\mathbf{a}_\mathrm{in}(\omega)\rangle \equiv \sqrt{\gamma} G(\omega) \langle\mathbf{a}_\mathrm{in}(\omega)\rangle$. In the last step, we introduced the system's Green's function $G(\omega) \equiv - \mathrm{i}(\omega - H)^{-1}$.

While the eigenvalues of $H$ are relevant for the time-evolution and dynamical stability of the system, the singular value decomposition discussed in the main text is relevant for the (steady state) system response which is experimentally accessible in a variety of platforms.
The response of a system typically depends on the frequency of the input field (as resonant and off-resonant response are expected to differ greatly) as well as gain and loss.
It has therefore been shown~\cite{Brunelli2023Restoration,Porras2019} that also a theory that connects non-Hermitian topology to different steady-state responses has to invoke a tool that provides a notion of distance from the origin.
While the eigendecomposition is invariant under diagonal shifts, the singular value decomposition fulfills this task.

When $H$ is non-normal $[H,H^\dagger]\neq0$, the singular value decomposition and the eigendecomposition can differ drastically in their behaviour. In particular, the eigendecomposition can display the non-Hermitian skin effect, while the singular value decomposition does not.
While the eigendecomposition prevents us from establishing a straightforward bulk-boundary correspondence, it was shown that for one-dimensional one-band systems, the bulk-boundary correspondence can be restored with the help of the singular value decomposition~\cite{Porras2019,Herviou2019,Brunelli2023Restoration}, relating non-trivial values of a non-Hermitian winding number to a corresponding number of localised zero modes.
These topological zero modes then dominate the scattering response~\cite{Porras2019,Brunelli2023Restoration}
\begin{align}\label{eq:ScatMatSVD}
    S(\omega)
    & \cong \mathbb{1} - \mathrm{i} \sqrt{\gamma} \sum_{j\in \mathcal{S}_\mathrm{ZM}} \frac{1}{\sigma_j} \ketbra{v_j}{u_j} \sqrt{\gamma}
\end{align}
with $\mathcal{S}_\mathrm{ZM}$ the set of zero modes.
The corresponding left and right singular vectors are localised at opposite ends, giving rise to directional end-to-end transport. The singular value sets the gain via $1/\sigma_j$ while the left singular vector selects the site producing the largest response and the right singular vector determines the output site.
Similarly, the steady state of Eq.~\eqref{eq:eoms} in NH non-trivial phases is governed by the zero mode
\begin{align}\label{eq:steadyState}
    \langle \mathbf{a}_\mathrm{ss} \rangle \cong -\mathrm{i}\sqrt{\gamma} \sum_{j\in \mathcal{S}_\mathrm{ZM}} \frac{1}{\sigma_j} \ketbra{v_j}{u_j} \langle \mathbf{a}_\mathrm{in}\rangle.
\end{align}
Since the SVD does not suffer from the NHSE, it is the perfect tool to study the interplay between non-Hermitian and Hermitian topology as shown in the main text.

\subsection*{Numerical computation of line- and point-gap invariants}

\noindent\textbf{Line gaps:}
For numerical purposes, it is convenient to define the band invariant $\mathcal{W}_j$ via a Wilson loop which is automatically gauge invariant
\begin{align}
    \mathcal{W}_j & = -\frac{1}{\pi}\mathrm{Im}\log\prod_\ell \left(\braket{ u_j(k_\ell)}{u_j(k_{\ell+1})}+\braket{v_j(k_\ell)}{v_j(k_{\ell+1})}\right)
\end{align}
in which $k_{\ell+1}$ is to be taken $\bmod 2\pi$. This expression recovers the invariant~\eqref{eq:ZakPhase} in the limit $k_{\ell+1}-k_\ell\to 0$.
Setting $k_\ell = \ell \Delta k$ with some sufficiently small $\Delta k$ provides a convenient formula for approximating the band invariant $\mathcal{W}_j$ from which we can compute the invariant of each line gap.

\noindent\textbf{Point gaps:}
The point gap invariant can conveniently be determined from a plot of the GSS in the complex plane. However, it is, in principle, also possible to compute the winding number via a Wilson loop
\begin{align}
    \nu_j & = -\frac{1}{2\pi}\mathrm{Im}\log\prod_\ell \left(\braket{ u_j(k_\ell)}{u_j(k_{\ell+1})}-\braket{v_j(k_\ell)}{v_j(k_{\ell+1})}\right).
\end{align}

\subsection*{Numerical computation of the generalised singular spectrum}

The numerical computation of the generalised singular spectrum requires some care. For instance, the naive approach of computing the SVD of $H(k)$ under PBC for each $k$ may result in the incorrect reconstruction of the phase $\phi_j\equiv \braket{u_j(k)}{v_j(k)}$ since the SVD for systems with additional symmetries (such as translational invariance) may not need to be not unique. Typically, the singular values are degenerate (each GSS band is periodic modulo $2\pi$), so within this degenerate subspace arbitrary superpositions of singular vectors are possible which distorts the reconstructed phase $\phi_j$. For the PBC case, one option to overcome this challenge is to solve for the SVD of $H(k)$ for a given $k$.
More generally, one can introduce a small fudge factor $\eta$ which breaks the symmetry giving rise to the degeneracy. Hence, we numerically solve for the eigenvectors of
\begin{align}
    \tilde{\mathcal{H}}
    \begin{pmatrix}
        \ket{\tilde v_j} \\ \ket{\tilde u_j}
    \end{pmatrix}
    = \begin{pmatrix}
        \eta \mathbb{1} & H^\dagger \\
        H & 0
    \end{pmatrix}
    \begin{pmatrix}
        \ket{\tilde v_j} \\ \ket{\tilde u_j}
    \end{pmatrix}
    = \tilde \sigma_j
    \begin{pmatrix}
        \ket{\tilde v_j} \\ \ket{\tilde u_j}
    \end{pmatrix}.
\end{align}
In the limit $\eta\to 0$ $\ket{\tilde v_j}$, $\ket{\tilde u_j}$ and $\tilde \sigma_j$ approach the correct left and right singular vectors and the singular values. In practice, we found that setting $\eta=10^{-9}$ yields to an excellent agreement between the numerically and the analytically computed GSS.
This approach can be applied both under PBC and OBC.
Under OBC, the boundary conditions need to be satisfied which typically leads to a hybridisation of multiple PBC singular vectors. As a result, the phase $\phi_j$ typically differs from the PBC phase, see Fig.~\ref{fig:edgeModes}, while the singular values with the exception of zero modes closely resemble the PBC singular values.

\subsection*{Further details about the two-dimensional non-Hermitian model with corner modes}

Here, we present further details about the physical couplings giving rise to the model~\eqref{eq:modelHOTI} as well as expand on the derivation of the topological invariants~\eqref{eq:windingGSSHOTI}.

\textbf{Physical model:}
We obtain the model~\eqref{eq:modelHOTI} through a combination of hopping with a complex hopping constant, non-local dissipation and incoherent pumping. This ensures that the doubled system, constructed according to Eq.~\eqref{eq:SVDdoubled}, corresponds to the Hermitian BBH model~\cite{Benalcazar2017Quantized,Benalcazar2017Electric}.
Specifically, $A$ and $B$ sub-lattices are each coupled along the $x$ direction through a combination of hopping $\sum_{j_x,j_y} J_x (a_{j_x,j_y}^\dagger a_{j_x+1,j_y} + b_{j_x,j_y}^\dagger b_{j_x+1,j_y}) + \mathrm{h.c.}$ and dissipative couplings with dissipators $\sum_{j_x,j_y} \Gamma (\mathcal{D}[a_{j_x,j_y} + e^{-\mathrm{i}\theta} a_{j_x+1,j_y}] + \mathcal{D}[b_{j_x,j_y} + e^{\mathrm{i}\theta} b_{j_x+1,j_y}])$, in which $a_{j_x,j_y}$ and $b_{j_x,j_y}$ denote the modes on the $A$ and $B$ sub-lattice at site $(j_x,j_y)$.
Tuning $J_x = \Gamma/2$ and $\theta=\pi/2$ results in perfect uni-directional coupling from left to right for the $A$ sub-lattice and from right to left for the $B$ sub-lattice, Fig.~\ref{fig:HOTI}.
Along the $y$ direction, imaginary intra-cell hopping of strength $g$ connects the $A$ to the $B$ sub-lattice according to $\sum_{j_x,j_y} \mathrm{i} g (a_{j_x,j_y} b_{j_x,j_y} - \mathrm{h.c.})$, while imaginary hopping of strength $J$ connects $A$ and $B$ sub-lattices between unit cells according to $\sum_{j_x,j_y} \mathrm{i} J (a_{j_x,j_y}^\dagger b_{j_x,j_y+1} + \mathrm{h.c.})$.
Additionally, each site experiences local losses at rate $\gamma$ according to $\sum_{j_x,j_y} \gamma (\mathcal{D}[a_{j_x,j_y}] + \mathcal{D}[b_{j_x,j_y}])$ and incoherent pumping at rate $\kappa$ according to $\sum_{j_x,j_y} \kappa (\mathcal{D}[a_{j_x,j_y}^\dagger] + \mathcal{D}[b_{j_x,j_y}^\dagger])$ to compensate some of the dissipation. Taken together, these decay rates give rise to the effective local dissipation rate $\gamma_\mathrm{eff} = \gamma+2\Gamma - \kappa$.

\textbf{Winding number characterisation:}
Hermitian higher-order topological insulators, such as the BBH model, can be characterised through winding numbers~\cite{Trifunovic2020Bulk,Ren2021Quadrupole,Yang2023Wannier} along $x$ and $y$, respectively. In Hermitian, fermionic systems, these winding numbers are associated with the respective edge polarisations while their product can be related to the quadrupole moment of the system. Even when only one edge, e.g. along $x$ yields a non-trivial winding number, this can result in corner states. This is the scenario we see in the main text in the non-Hermitian system corresponding to the BBH model.

\begin{figure}
    \centering
    \includegraphics[width=\linewidth]{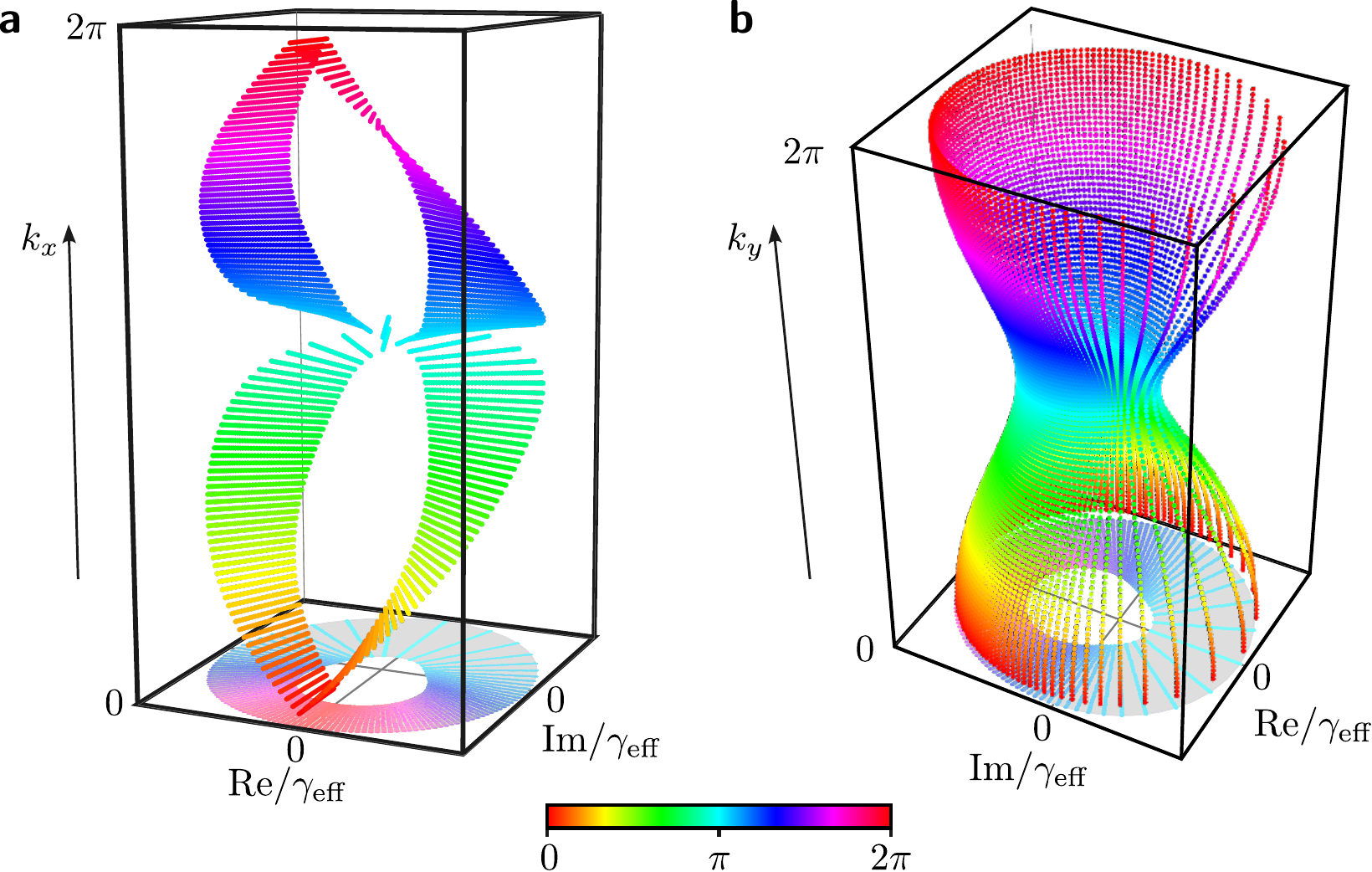}
    \caption{\textbf{Generalised singular spectrum as a function of quasi-momentum.}
    Plotting the generalised singular spectrum (GSS) in a 3D plot with \textbf{a}~$k_x$ or \textbf{b}~$k_y$ on one of the axes reveals the two bands and the orientation of the windings of each band. As $k_x$ is varied, each band winds, in opposite directions, around the origin.
    Here, $g/\gamma_\mathrm{eff}=0.6$, $J/\gamma_\mathrm{eff}=1$, $\Gamma/\gamma_\mathrm{eff}=0.6$, $N=100$.
    }
    \label{fig:NHHOTI3D}
\end{figure}

In analogy to these Hermitian higher-order topological insulators, we can compute the winding numbers of our non-Hermitian system according to
\begin{align}
    \nu_{\alpha}(k_\beta) = \frac{1}{2\pi\mathrm{i}} \int_0^{2\pi} \mathrm{d} k_\alpha \,
    \partial_{k_\alpha} \ln \det H(k_x,k_y)
\end{align}
for a fixed $k_\beta$ with $\alpha\in\{x,y\}$ and $\beta$ the respective other label from $\{x,y\}$.
To obtain the winding number of the $j$th band along direction $\alpha$, we also integrate along the perpendicular $k_\beta$
\begin{align}
    \nu_{\alpha}
    & = \frac{1}{2\pi} \int_0^{2\pi} \mathrm{d}k_\beta \, \nu_{j,\alpha}(k_\beta) \notag\\
    & = \frac{1}{(2\pi)^2\mathrm{i}} \iint_0^{2\pi} \mathrm{d} k_\beta \mathrm{d} k_\alpha \,
    \partial_{k_\alpha} \ln \det H(k_x,k_y)
    \label{eq:totalWindingHOTI}
\end{align}
Since $\nu_{j,\alpha}(k_\beta)$ is constant over $k_\beta$, we have to divide by another factor of $2\pi$.
The connection between this winding number and Eq.~\eqref{eq:windingGSSHOTI} involving the singular vectors follows in an analogous way to the 1D case (see the SI for the derivation in the 1D case).
Using $\partial_{k_\alpha} \ln \det H(k_x,k_y) = \mathrm{tr}\, \partial_{k_\alpha} \ln H(k_x,k_y) = \mathrm{tr}\, H^{-1}(k_x,k_y) \partial_{k_\alpha} H(k_x,k_y)$ and expanding $H(k_x,k_y)$ into the singular values and vectors, we can split the integral~\eqref{eq:totalWindingHOTI} into integrals for the individual bands $\nu_{\alpha} = \sum_j \nu_{j,\alpha}$ with $\nu_{j,\alpha}$
\begin{align}
    \nu_{j,\alpha} = \frac{1}{(2\pi)^2\mathrm{i}} \iint_0^{2\pi} \mathrm{d}k_\alpha \mathrm{d}k_\beta \,
    (
    &
    \bra{u_j(k_x,k_y)} \partial_{k_\alpha} \ket{u_j(k_x,k_y)} \notag \\
    - &
    \bra{v_j(k_x,k_y)}\partial_{k_\alpha} \ket{v_j(k_x,k_y)}
    ).
\end{align}
Just as in the 1D case, this integral can be related to an integral over the relative phase between left and right singular vector $\mathrm{Arg}\,\braket{u_j(k_x,k_y)}{v_j(k_x,k_y)}$ and we obtain Eq.~\eqref{eq:windingGSSHOTI} of the main text.
As discussed in the main text and illustrated in Fig.~\ref{fig:NHHOTI3D}, the two bands wind as a function of $k_x$, so $\nu_{j,x} = \pm 1$, while $\nu_{j,y}=0$.

\setcounter{figure}{0}
\renewcommand{\thefigure}{S\arabic{figure}}
\renewcommand{\theequation}{S\arabic{equation}}
\setcounter{equation}{0}
\setcounter{section}{0}
\setcounter{table}{0}
\clearpage

\onecolumngrid

\begin{center}
    {\large\bf Supplementary Information for ``Unifying framework for non-Hermitian and Hermitian topology in driven-dissipative systems''} \\[2ex]
	Clara C. Wanjura, Andreas Nunnenkamp
\end{center}

\section{Derivation of the winding number formula}
In the main text, we presented the formula for the winding number associated with a point gap
\begin{align}
    \nu_j = \frac{1}{2\pi\mathrm{i}} \int_0^{2\pi}\mathrm{d}k\,\partial_k\mathrm{Arg}\,\braket{u_j}{v_j} = \frac{1}{2\pi\mathrm{i}} \int_0^{2\pi}\mathrm{d}k\,\frac{\partial_k\braket{u_j}{v_j}}{\braket{u_j}{v_j}},
\end{align}
Here, we derive this formula.
As described in the main text, the total winding number of the system is given by
\begin{align}
    \nu & =
    \frac{1}{2\pi\mathrm{i}} \int_0^{2\pi} \mathrm{d}k\, \mathrm{tr}\,\left(\partial_k \ln H\right) \equiv \sum_j \nu_j
\end{align}
Using the SVD, one can write
\begin{align}
    \partial_k \ln H = H^{-1} \partial_k H = \sum_{j,\ell} \frac{1}{\sigma_j(k)}\ketbra{v_j(k)}{u_j(k)} \left(\partial_k \sigma_\ell(k) \ketbra{u_\ell(k)}{v_\ell(k)} \right)
\end{align}
and
\begin{align}
    \mathrm{tr}\partial_k \ln H & = \sum_m \bra{v_m(k)} \sum_{j,\ell} \frac{1}{\sigma_j(k)}\ketbra{v_j(k)}{u_j(k)} \left(\partial_k \sigma_\ell(k) \ketbra{u_\ell(k)}{v_\ell(k)} \right) \ket{v_m(k)} \\
    & = \sum_{j,\ell}
    \frac{\sigma_\ell(k)}{\sigma_j(k)}
    \bra{u_j(k)} \left(\partial_k \ketbra{u_\ell(k)}{v_\ell(k)} \right) \ket{v_j(k)}
    +
    \sum_j \frac{\partial_k \sigma_j(k)}{\sigma_j(k)} \\
    & = \sum_{j,\ell} \frac{\sigma_\ell(k)}{\sigma_j(k)} \big(
    \bra{u_j(k)} \partial_k \ket{u_\ell(k)}
    \underbrace{\braket{v_\ell(k)}{v_j(k)}}_{=\delta_{j,\ell}}
    +
    \underbrace{\braket{u_j(k)}{u_\ell(k)}}_{=\delta_{j,\ell}}
    \left(\partial_k\bra{v_\ell(k)}\right) \ket{v_j(k)}
    \big)
    +
    \sum_j \frac{\partial_k \sigma_j(k)}{\sigma_j(k)} \\
    & = \sum_{j} \left(
    \bra{u_j(k)} \partial_k \ket{u_j(k)}
    +
    \left[\partial_k\bra{v_j(k)}\right] \ket{v_j(k)}
    \right)
    +
    \underbrace{\sum_j \frac{\partial_k \sigma_j(k)}{\sigma_j(k)}}_{\text{winding number zero}}
\end{align}
in which we used the fact that the $\ket{v_j(k)}$ and $\ket{u_j(k)}$ each span an orthonormal basis $\braket{v_j(k)}{v_\ell(k)}=\delta_{j,\ell}$, $\braket{u_j(k)}{u_\ell(k)}=\delta_{j,\ell}$ to compute the trace. Since the singular values are non-negative, they cannot wind around the origin, so $\partial_k \sigma_j(k)/\sigma_j(k)$ has winding number zero
and we can neglect this term in the overall computation of the winding number
\begin{align}
    \nu & = \frac{1}{2\pi\mathrm{i}} \int_0^{2\pi} \mathrm{d}k\, \mathrm{tr}\,\left(\partial_k \ln H\right) \\
    & = \frac{1}{2\pi\mathrm{i}} \int_0^{2\pi} \mathrm{d}k\,
    \sum_{j} \left(
    \bra{u_j(k)} \partial_k \ket{u_j(k)}
    +
    [\bra{v_j(k)}\partial_k \ket{v_j(k)}]^*
    \right) \\
    & = \frac{1}{2\pi\mathrm{i}} \int_0^{2\pi} \mathrm{d}k\,
    \sum_{j} \left(
    \bra{u_j(k)} \partial_k \ket{u_j(k)}
    -
    \bra{v_j(k)}\partial_k \ket{v_j(k)}
    \right).
\end{align}
In the last step, we used that $[\bra{v_j(k)}\partial_k \ket{v_j(k)}]^* = - \bra{v_j(k)}\partial_k \ket{v_j(k)}$ which holds because
\begin{align}
    0 = \partial_k \delta_{j,\ell} = \partial_k \braket{v_j(k)}{v_j(k)} = [\partial_k \bra{v_j(k)}] \ket{v_j(k)} + \bra{v_j(k)}\partial_k \ket{v_j(k)} = [\bra{v_j(k)}\partial_k \ket{v_j(k)}]^* + \bra{v_j(k)}\partial_k \ket{v_j(k)}.
\end{align}
We can therefore write the winding number as sum over the winding numbers $\nu_j$ of individual bands defined by the left and right singular vectors
\begin{align}
    \nu
    & = \frac{1}{2\pi\mathrm{i}} \sum_{j}  \int_0^{2\pi} \mathrm{d}k\,
    \left(
    \bra{u_j(k)} \partial_k \ket{u_j(k)}
    -
    \bra{v_j(k)}\partial_k \ket{v_j(k)}
    \right) \equiv \sum_j \nu_j
\end{align}
with
\begin{align}
    \nu_j = \frac{1}{2\pi\mathrm{i}}  \int_0^{2\pi} \mathrm{d}k\,
    \left(
    \bra{u_j(k)} \partial_k \ket{u_j(k)}
    -
    \bra{v_j(k)}\partial_k \ket{v_j(k)}
    \right).
\end{align}

\noindent\textbf{Connection to the phase:}
Partitioning the Brillouin zone into segments of length $\Delta k$ and introducing $k_j\equiv k_{j-1}+\Delta k$ with some $k_0\in[0,2\pi)$, this expression can be recast into the form
\begin{align}
    \nu_j & = \frac{1}{2\pi\mathrm{i}}\int_0^{2\pi}\mathrm{d}k\,
    \left(\bra{u_j(k)} \partial_k \ket{u_j(k)}
    -
    \bra{v_j(k)}\partial_k \ket{v_j(k)}\right)
    = \frac{1}{2\pi}\lim_{\Delta k\to 0}\left(\mathrm{Arg}\prod_\ell \braket{u_j(k_\ell)}{u_j(k_{\ell+1})} - \mathrm{Arg}\prod_\ell\braket{v_j(k_\ell)}{v_j(k_{\ell+1})}\right).
\end{align}
Intuitively, one of the terms on the right-hand side is the accumulation of phases as we parallel transport $\ket{u_j(k)}$ and $\ket{v_j(k)}$ along the path defined by the integral on the left-hand side.
With $\braket{u_j(k_\ell)}{u_j(k_{\ell+1})} = e^{\mathrm{i}\Delta\phi_{j,\ell}^{u}}$ and $\braket{v_j(k_\ell)}{v_j(k_{\ell+1})} = e^{\mathrm{i}\Delta\phi_{j,\ell}^{v}}$, we obtain
\begin{align}
    \nu_j & = \frac{1}{2\pi\mathrm{i}}\int_0^{2\pi}\mathrm{d}k\,
    \left(\bra{u_j(k)} \partial_k \ket{u_j(k)}
    -
    \bra{v_j(k)}\partial_k \ket{v_j(k)}\right)
    = \frac{1}{2\pi}\lim_{\Delta k\to 0} \sum_\ell \left(\Delta\phi_{j,\ell}^{v} - \Delta\phi_{j,\ell}^{u}\right) \\
    & = \frac{1}{2\pi\mathrm{i}}\int_0^{2\pi}\mathrm{d}k\, \partial_k\mathrm{Arg}\,\braket{u_j}{v_j}
    = \frac{1}{2\pi\mathrm{i}}\int_0^{2\pi}\mathrm{d}k\,\frac{\partial_k\braket{u_j}{v_j}}{\braket{u_j}{v_j}}.
\end{align}

Note that alternatively, we can derive this formula by rewriting the SVD into a polar-decomposition $H(k) = U(k) \Sigma(k) V^\dagger(k) = U(k) V^\dagger(k) V(k) \Sigma(k) V^\dagger \equiv W(k) R(k)$ with $W(k) \equiv U(k) V^\dagger(k)$ and $R(k) \equiv V(k) \Sigma(k) V^\dagger$. The total winding number of $\mathrm{det}\,H(k)$ is then determined by the trace of $W(k) = U(k) V^\dagger(k)$ which is exactly the sum of the winding numbers defined above.

\section{Symmetries of the NH SSH model}
Here, we discuss the symmetries of the NH SSH model, Eq.~\eqref{eq:BlochMatNHSSH} of the main text, on the level of the doubled matrix, Eq.~\eqref{eq:SVDdoubled} of the main text. 
First, we rewrite $H(k)$, Eq.~\eqref{eq:BlochMatNHSSH} of the main text, as
\begin{align}
    H(k) & = \begin{pmatrix*}
    - \mathrm{i}\gamma_\mathrm{eff}/2 & J_{L1} + J_{L2} e^{\mathrm{i} k} \\
    J_{R1} + J_{R2} e^{-\mathrm{i} k} & - \mathrm{i}\gamma_\mathrm{eff}/2
    \end{pmatrix*} \notag \\
    & = - \mathrm{i}\frac{\gamma_\mathrm{eff}}{2} \mathbb{1} + (J_{L1} + J_{L2} e^{\mathrm{i} k}) \frac{\sigma_x + \mathrm{i}\sigma_y}{2} + (J_{R1} + J_{R2} e^{-\mathrm{i} k}) \frac{\sigma_x - \mathrm{i}\sigma_y}{2}
\end{align}
with
$J_{L1}\equiv J_1-\mathrm{i}\frac{\Gamma_1}{2}e^{-\mathrm{i}\theta_1}$,
$J_{L2}\equiv J_2-\mathrm{i}\frac{\Gamma_2}{2}e^{\mathrm{i}\theta_2}$,
$J_{R1}\equiv J_1-\mathrm{i}\frac{\Gamma_1}{2}e^{\mathrm{i}\theta_1}$,
$J_{R2}\equiv J_2-\mathrm{i}\frac{\Gamma_2}{2}e^{-\mathrm{i}\theta_2}$.
The corresponding doubled Hermitian matrix is then given by
\begin{align}\label{eq:doubledHamiltonian}
    \mathcal{H}(k) & = \begin{pmatrix}
        0 & H^\dagger(k) \\
        H(k) & 0
    \end{pmatrix} \notag\\
    & = \begin{pmatrix}
        0 & 0 & \mathrm{i}\gamma_\mathrm{eff}/2  & J_{R1} + J_{R2} e^{\mathrm{i} k} \\
        0 & 0 & J_{L1} + J_{L2} e^{-\mathrm{i} k}  & \mathrm{i}\gamma_\mathrm{eff}/2 \\
        - \mathrm{i}\gamma_\mathrm{eff}/2 & J_{L1} + J_{L2} e^{\mathrm{i} k} & 0 & 0 \\
        J_{R1} + J_{R2} e^{-\mathrm{i} k} & - \mathrm{i}\gamma_\mathrm{eff}/2 & 0 & 0
    \end{pmatrix}
\end{align}
Apart from chiral symmetry which the model preserves by construction, the model has additional symmetries, when $J_{Lj}$, $J_{Rj}$ are real which is automatically the case when $\theta=\pm\frac{\pi}{2}$.
For the discussion of the Hermitian--like topological edge modes, we only required particle-hole symmetry to ensure the quantisation of the invariant $\mathcal{W}$. Beyond that, the doubled Hermitian matrix of the NH SSH model displays time-reversal symmetry, inversion symmetry and can also preserve PT symmetry as we demonstrate below.

\noindent\textbf{Particle-hole symmetry and time-reversal symmetry:}
It is straightforward to check that $\mathcal{H}(k)$ fulfills $\mathcal{U}^\dagger \mathcal{H}^*(-k) \mathcal{U} = - \mathcal{H}(k)$ with $\mathcal{U} \equiv \mathrm{diag}(-\mathrm{i},\mathrm{i},-\mathrm{i},\mathrm{i})$. Since the model also preserves chiral symmetry by construction, i.e., $\mathcal{R}^\dagger \mathcal{H}(k) \mathcal{R} = - \mathcal{H}(k)$ 
with $\mathcal{R} = \sigma_z \otimes \sigma_z$, the model also automatically obeys time-reversal symmetry.

\noindent\textbf{Inversion symmetry:}
As can be easily checked, we have $\mathcal{V}^\dagger\mathcal{H}(-k) \mathcal{V} = - \mathcal{H}(k)$. Specifically, we define $\mathcal{V}\equiv \mathcal{U} \mathcal{J} \mathcal{U}^\dagger$ with the gauge transformation $\mathcal{U}=\mathrm{diag}(e^{\mathrm{i}\pi/4},e^{-\mathrm{i}\pi/4},e^{\mathrm{i}\pi/4},e^{-\mathrm{i}\pi/4})$ and the exchange matrix $\mathcal{J}$
\begin{align}
    \mathcal{J} & = \begin{pmatrix*}
        0 & 0 & 0 & 1 \\
        0 & 0 & 1 & 0 \\
        0 & 1 & 0 & 0 \\
        1 & 0 & 0 & 0
    \end{pmatrix*} = \sigma_x \otimes \sigma_x
\end{align}
which here performs the role of transposition.
The intuition between this transformation is that the gauge transformation first distributes the $\pi/2$ phase stemming from the dissipation term evenly within the unit cell and the exchange matrix transposes the matrix which together with $k\to-k$ plays the role of a spatial reflection.
Taken together, we have $\mathcal{V}=\sigma_x\otimes\sigma_y$.

\noindent\textbf{Real gauge and PT symmetry:}
$\mathcal{H}(k)$ obeys PT symmetry when it can be transformed to a purely real matrix.
Specifically, $\mathcal{H}(k)$ is real in the basis defined by the unitary
\begin{align}
        \mathcal{S} \equiv \frac{1}{\sqrt{2}}
            \begin{pmatrix*}
            1 & 0 & 0 & 1 \\
            -\mathrm{i} & 0 & 0 & \mathrm{i} \\
            0 & 1 & 1 & 0 \\
            0 & -\mathrm{i} & \mathrm{i} & 0
        \end{pmatrix*}.
\end{align}
Concretely,
\begin{align}
    \mathcal{S}^\dagger \mathcal{H}(k) \mathcal{S} & =
    \begin{pmatrix*}
    J_{R1} + J_{R2} \cos k &  J_{R2} \sin k & 0 & -\gamma/2 \\
    J_{R2} \sin k & -J_{R1} - J_{R2} \cos k  & -\gamma/2 & 0 \\
    0 & -\gamma/2 & J_{L1} + J_{L2} \cos k & - J_{L2} \sin k \\
    -\gamma/2 & 0 & - J_{L2} \sin k & -J_{L1} - J_{L2} \cos k 
    \end{pmatrix*}.
\end{align}

\section{Trivial edge modes}
\begin{figure*}
    \includegraphics[width=\textwidth]{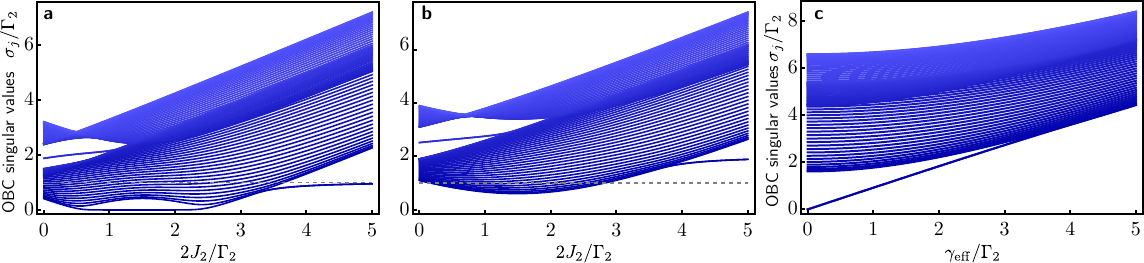}
    \caption{\textbf{Trivial edge modes can be absorbed into the bulk without closing a gap.}
    Edge modes that we can understand as remnants of Hermitian topology but are not protected by any symmetry can be absorbed into the bulk simply by increasing the local dissipation without closing any gap. The grey line in \textbf{a} and \textbf{b} indicates the asymptotic value of the singular values associated with the edge modes in panel \textbf{a}.
    Here, \textbf{a}~$\gamma_\mathrm{eff}/\Gamma_2=1$,
    \textbf{b}~$\gamma_\mathrm{eff}/\Gamma_2=2$;
    \textbf{a}-\textbf{c}~$\theta=\frac{\pi}{2}$, $J_2/\Gamma_2 = J_1/\Gamma_2=1.5$, $\Gamma_1/\Gamma_2=0.4$, $N=40$.}
    \label{fig:HermitianEdgeModesAbsorbed}
\end{figure*}%
Edge modes that we can understand as remnants of Hermitian topology but are not protected by any symmetry can be absorbed into the bulk simply by increasing the local dissipation without closing any gap. Here, we illustrate this for the NH SSH model of the main text. By increasing the local dissipation, we can remove the Hermitian-like edge modes without ever closing a gap. This is illustrated in Fig.~\ref{fig:HermitianEdgeModesAbsorbed}. Panel~\textbf{a} displays the sweep of the main text in which we indicate the asymptotic value of the edge modes as grey dashed line. Increasing the local dissipation $\gamma_\mathrm{eff}$ in panel~\textbf{b} elevates the edge modes to a higher singular value while the edge modes are already absorbed into the bulk for certain values of $2J_2/\gamma_\mathrm{eff}$. We also notice that the increase in dissipation has removed the NH edge mode through the closing of the point gap. NH edge modes are automatically protected by the chiral symmetry arising in the construction of the SVD so they can only be removed through point gap closings if present.
In panel~\textbf{c}, we fix $2J_2/\Gamma_2=4.5$ and sweep $\gamma_\mathrm{eff}/\Gamma_2$. Eventually, the edge modes are absorbed into the bulk.
This shows that, indeed, the Hermitian-like edge modes that are not protected by additional symmetries are trivial and can be removed without ever closing a gap.

The only way, Hermitian-like edge modes can lie within a band gap and be removed via gap closing in the GSS is on the onset of a dynamic instability.
The two bands then overlap but the degeneracy is resolved in the GSS. The edge mode is then removed via a line gap closing in the GSS passing through the origin.


\end{document}